\RequirePackage{fix-cm}
\documentclass[smallextended]{svjour3}       
\usepackage[utf8]{inputenc}
\usepackage{xspace}
\usepackage{tikz-cd}
\usepackage{ifsym}
\usepackage{comment}
\usepackage{ifthen}
\usepackage{titlesec}
\usepackage{enumitem}
\usetikzlibrary{positioning}

\titleformat{\paragraph}
{\normalfont\normalsize\bfseries}
{\theparagraph}{1em}{}

\usepackage{bbold}

\usepackage{graphicx}
\usepackage{mathptmx}      
\journalname{Journal of Statistical Physics}
\usepackage{xcolor}
\usepackage{amsmath}
\usepackage{amssymb}
\usepackage{mathtools}
\usepackage{hyperref}
\usepackage[caption=false]{subfig}

\newcommand{\REF}[2][]{
	\ifthenelse{\equal {#1} {}}{Ref.~\cite{#2}}{Ref.~\cite[#1]{#2}}}
\newcommand\subcap[1]{{(#1):}}
\newcommand\subfig[2]{{Fig.~\ref{#1}{#2}}}

\newcommand{\SET}[1]{\{#1\}}

\newcommand{\eq}[1]{eq.~(\ref{#1})}

\newcommand{\Eqq}[1]{Equation~(\ref{#1})}
\newcommand{\eqtwo}[2]{eqs.~(\ref{#1}) and~(\ref{#2})}
\newcommand{\eqfromto}[2]{eq.~(\ref{#1}) to eq.~(\ref{#2})}

\newcommand{\fig}[1]{Fig.~\ref{#1}}
\newcommand{\quot}[1]{``#1''}

\newcommand{\app}[1]{Appendix~\ref{#1}}
\newcommand{\sect}[1]{Sec.~\ref{#1}}

\newcommand{\etc}{\textrm{etc.}}
\newcommand{\etcp}{\textrm{etc}}

\newcommand{\Tr}{\text{Tr}}  

\newcommand{\OCAL}{\mathcal{O}}  
\newcommand{\epsilontilde}{\tilde{\epsilon}}

\newcommand{\BRA}[1]{\ensuremath{\langle #1 |}} 
\newcommand{\BRAN}[1]{\ensuremath{\langle #1 }} 
\newcommand{\KET}[1]{\ensuremath{|#1 \rangle }} 
\newcommand{\expb}[1]{\exp \glb #1 \grb} 
\newcommand{\sina}[2][]{\sin^{#1} \! \gla #2 \gra}  
\newcommand{\cosa}[2][]{\cos^{#1} \! \gla #2 \gra}  

\newcommand{\sinb}[2][]{\sin^{#1} \glb #2 \grb}  
\newcommand{\cosb}[2][]{\cos^{#1} \glb #2 \grb}  

\newcommand{\sinc}[2][]{\sin^{#1} \glc #2 \grc}  
\newcommand{\cosc}[2][]{\cos^{#1} \glc #2 \grc}  

\newcommand{\minb}[2][]{\min^{#1} \glb #2 \grb}  

\newcommand{\R}{\mathbb{R}}

\newcommand{\gla}{\,}  
\newcommand{\gra}{}  
\newcommand{\glb}{\left(}  
\newcommand{\grb}{\right)}  
\newcommand{\glc}{\left[}  
\newcommand{\grc}{\right]}  
\newcommand{\gld}{\left\{}  
\newcommand{\grd}{\right\}}  
\newcommand{\gle}{\left|}  
\newcommand{\gre}{\right|}  

\newcommand{\const}{\text{const}}

\newcommand{\PLUSPLUS}{+ \dots +}

\newcommand{\TO}{,\ldots,}

\newcommand{\half}{\frac{1}{2}}
\newcommand{\thalf}{\tfrac{1}{2}}
\newcommand{\quarter}{\frac{1}{4}}

\newcommand\bigOb[1]{\ensuremath{\OCAL\glb #1 \grb}}

\newcommand{\taurel}{\tau_{\text{rel}}}
\newcommand{\squarewave}{\text{square}}
\newcommand{\flatshape}{\text{flat}}
\newcommand{\wedgeshape}{\text{wedge}}
\newcommand{\Vshape}{\text{V}}
\newcommand{\tmix}[1][]{t_{\text{mix}}}
\newcommand{\pit}[1]{\pi^{\{#1\}}}
\newcommand{\taukemeny}{\tau^{*}}
\newcommand{\Kemeny}{Kemeny time\xspace}
\newcommand{\Kemenys}{Kemeny times\xspace}
\newcommand{\MFPT}{mean first-passage time\xspace}
\newcommand{\MFPTs}{mean first-passage times\xspace}
\newcommand{\MFPTCAP}{Mean first-passage time\xspace}
\newcommand{\WEDGE}{wedge\xspace}
\newcommand{\WEDGECAP}{Wedge\xspace}
\newcommand{\V}{V-shape\xspace}
\newcommand{\minfunc}[2]{\min\glb \! 1, \tfrac{\pi_{#1}}{\pi_{#2}} \!\grb }
\newcommand{\minfunctext}[2]{\min\glb \! 1, \pi_{#1}/ \pi_{#2} \!\grb }
\renewcommand{\R}[1]{r^{\{ #1\}}}
\renewcommand{\L}[1]{l^{\{ #1\}}}
\newcommand{\ONE}{\mathbb{1}}
\newcommand{\ZERO}{\mathbb{0}}
\newcommand{\ONEPI}{\KET{1}\BRA{\pi}}
\newcommand{\RMLM}{\KET{\R{m}}\BRA{\L{m}}}
\newcommand{\DiracCoord}{\, \, \Longleftrightarrow \, \,}

\newcommand{\LIFTED}[2]{#1^{#2}}
\newcommand{\ExactDiag}{explicit numerical matrix diagonalizations\xspace}
\newcommand{\ExactDiagShort}{matrix diagonalizations\xspace}
\newcommand{\GM}{Green's matrix\xspace}
\newcommand{\GUM}{Green's-matrix\xspace}
\newcommand*\circled[1]{\tikz[baseline=(char.base)]{
    \node[shape=circle, draw, inner sep=1pt,
        minimum height=12pt] (char) {#1};}}

\newcommand{\oo}{\cdot}
\newcommand{\Gcal}[2]{G^{#1,\sigma''}_{#2 k}}
\newcommand{\glift}[1]{g^{[k, \sigma'']}_{#1}}

\newcommand{\colored}{} 
\ifdefined\colored
	
\else
	
\fi

\begin{document}

\title{Markov chains at the onset of non-reversibility
}


\author{Gustave Robichon \and Cécile Monthus \and Werner Krauth
}


\institute{
Gustave Robichon\\
Laboratoire de Physique de l’Ecole Normale Supérieure - PSL, \\
Centre Automatique et Systèmes Mines Paris - PSL, \\
CNRS, Inria, PSL Research University, Paris, France \\
\email{gustave.robichon23@gmail.com}\\
\and
Cécile Monthus\\
Université Paris-Saclay, CNRS, CEA, Institut de
Physique théorique\\
91191  Gif-sur-Yvette, France\\
\email{cecile.monthus@ipht.fr} \\
\and
Werner Krauth \\
Laboratoire de Physique de l'Ecole normale supérieure, ENS,
Université PSL, CNRS, Sorbonne Université, Université de Paris Cité,
Paris, France\\
Rudolf Peierls Centre for Theoretical Physics \\ University
of Oxford (UK)
\email{werner.krauth@ens.fr}\\
}

\date{Received: date / Accepted: date}

\maketitle
\setcounter{tocdepth}{3}
\tableofcontents

\begin{abstract}
For a one-dimensional path graph and a lifted path graph constructed from a
duplication of each of its sites, we study how a
reversible Markov chain can be perturbed and gradually driven into
non-reversibility. The reversible Markov chain has a transition matrix that is
diagonalizable and features
real-valued eigenvalues and eigenvectors. The left and right
eigenvectors form a biorthogonal system.
We discuss in concrete examples how the transition matrix of a
non-reversible Markov
chain may be diagonalizable or non-diago\-nalizable, and it may have real
eigenvalues and complex-conjugate pairs.
For a number of steady states (flat, square-wave,
wedge, V-shape), we compute eigenvalue spectra on both graphs and discuss the
speedup that can be achieved through lifting. We develop a \GM formalism, which
we use to compute \Kemenys and \MFPTs, and which provides
valuable information
and allows us to interpret the results for the characteristic times.

\end{abstract}

\section{Introduction}
\label{sec:Introduction}

Non-reversible Markov chains are random processes that violate the
detailed-balance condition and, thus, time-reversal symmetry~\cite{Levin2008}. A
non-reversible Markov chain usually converges towards a steady state (in
the physics literature often
called \quot{NESS} for \quot{non-equilibrium steady state}), that is difficult
to characterize~\cite{Taniguchi2006}. Alternatively, a non-reversible Markov
chain
can also be designed to converge to
an imposed equilibrium distribution, for example the Boltzmann distribution.
This creates a  somewhat paradoxical situation, where a non-reversible (that is,
in physicists' terminology,
non-equilibrium) Markov chain converges towards a NESS that is carefully crafted
to be nothing but the equilibrium distribution. The design is motivated by the
fact that the non-reversible chain may converge much faster towards
\quot{equilibrium} than the equilibrium one:
The reversible
Markov chain, by definition has no steady-state flows, which generically
results in a diffusive process. In contrast, a non-reversible Markov chain can
feature ballistic motion even in the steady state.
In recent years, non-reversible
Markov chains have attracted widespread
attention~\cite{Diaconis2000,ChenLovaszPak1999,SuwaTodoPRL2010,Bernard2009}
\cite{Turitsyn2011,FernandesWeigelCPC2011,Krauth2021eventchain} as a means of
speeding up Monte Carlo algorithms for sampling an imposed equilibrium state,
with probability flows continuing even in the long-time limit.

Any reversible Markov chain has a diagonalizable transition matrix
with
real-valued eigenvalues and eigenvectors, and its left and right
eigenvectors form a bi\-orthogonal system~\cite[Sect.
12.1]{Levin2008}. In contrast, the transition matrix of a non-reversible Markov
chain may be diagonalizable or non-diago\-nalizable, and it may have only real
or real eigenvalues and complex-conjugate pairs. In spite of these
qualitative
differences between
the two classes of Markov chains, we can perturb a reversible Markov chain and
drive it gradually into non-reversibility. The qualitative differences induced
by non-reversibility then build up gradually. This is what we discuss in the
present paper.

\begin{figure}[htb]
	\centering
\includegraphics[width=\columnwidth]{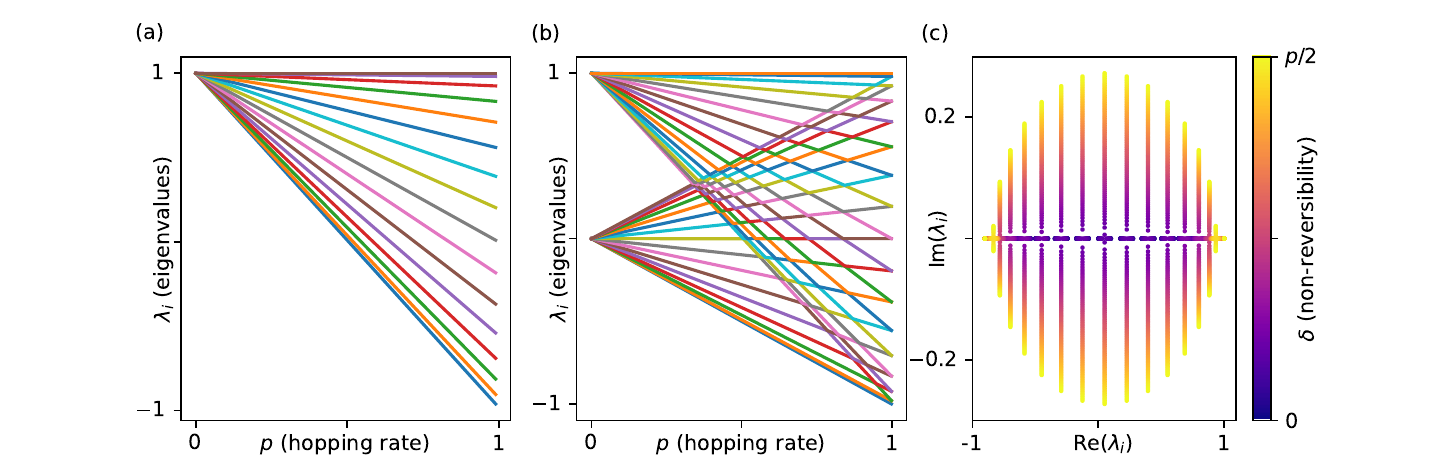}
  \caption{Transition-matrix spectra for the flat steady state
($N=16$).
  \subcap{a} Path graph of
\eq{equ:TransitionMatrixCollapsed} (reversible Markov chain).
  \subcap{b} Lifted path graph of
\eq{equ:LiftedTransitionReversible} (reversible Markov chain).
  \subcap{c} Lifted path graph of
\eq{equ:LiftedTransitionNonReversible} with $p=0.9$ as a
function of the non-reversibility $\delta$ (see also
\fig{fig:ReversibleNonReversible}). The complex spectrum arises through the
hybridization of the two fans in (b).
}
\label{fig:Flat_A_B_C}
\end{figure}

To illustrate in a simple example how a reversible Markov chain can be tuned
into non-reversibility and how non-reversibility can produce speedups, we
proceed in three
steps, closely following \REF{Diaconis2000}. In the first step, we consider a
particle performing a random walk on an $N$-site path
graph (without periodic boundary conditions) with,  for
$N=4$, the following hopping probabilities:
\begin{equation}
  \begin{tikzcd}[column sep = 1.2cm]
 \arrow[loop above]{}{\small{\underbrace{1 - \thalf p }}}
\circled{1} \arrow[yshift = 0.5 ex]{r}{\half p }&
 \arrow[loop above]{}{\small{\underbrace{1 - p}}}
\circled{2}
\arrow[yshift = -0.5 ex]{l} { \half p}
\arrow[yshift = 0.5 ex]{r}{\half p  }&
 \arrow[loop above]{}{\small{\underbrace{1- p }}}
\circled{3}
\arrow[yshift = 0.5 ex]{r}{\half p}
\arrow[yshift = - 0.5 ex]{l}{ \half p}
&
 \arrow[loop above]{}{\small{\underbrace{1 - \thalf p }}}
\circled{4}
\arrow[yshift = -0.5 ex]{l} { \half p }
\end{tikzcd}.
\label{equ:TransitionMatrixCollapsed}
\end{equation}
For transition rates $0<p \le 1$, in the limit $t \to
\infty$, the reversible Markov chain encoded in
\eq{equ:TransitionMatrixCollapsed} converges towards the flat steady state
$\SET{\pi_1 \TO  \pi_N} = \SET{1\TO 1}/N$. The transition matrix
satisfies the detailed-balance condition $\pi_i P_{ij} = \pi_j P_{ji}\quad
\forall i,j \in \Omega$, that is, it is reversible. On the path graph, detailed
balance reduces to
\begin{equation}
    \pi_{i} P_{i,i+1}    = \pi_{i+1} P_{i+1,i} \quad \forall i \in {1 \TO N-1}.
\label{equ:DetailedBalancePath}
\end{equation}
The eigenvalues of \eq{equ:TransitionMatrixCollapsed} are obtained by
diagonalizing the transition matrix using the Fourier basis, yielding:
\begin{equation}
  \lambda(h) = (1-p) + p \cosb{\pi\frac{h}{N}}, \quad h=0,1,\ldots,N-1,
\label{equ:SpectrumCollapsed}
\end{equation}
where we notice that $\lambda(0) = 1$
(see \subfig{fig:Flat_A_B_C}{a}).
For $p>0$, the eigenvalue $1$ is
non-degenerate, as
for any irreducible transition matrix, and there is no other eigenvalue of unit
norm, as for any aperiodic transition matrix. The corresponding left
eigenvector of $P$ is the steady state.
The entire spectrum is real-valued, as for any reversible transition
matrix.
As a function of the transition rate $p$, the spectrum forms a fan starting
at $1$ (for $p=0$), and for
fixed $0<p \le 1$, the transition matrix has
a large number of eigenvalues (those with $h \ll N$), that differ from $1$ by
$p \pi^2 h^2 / (2 N^2) \sim 1/N^2$.
We may consider the eigenvalues $\lambda$ of $P$  and define
\begin{equation}
 \lambda_* = \max \gld |\lambda| : \text{$\lambda$ is an eigenvalue of $P$,
$\lambda \neq 1$} \grd.
\label{equ:MaxAbs}
\end{equation}
The gap of the spectrum (also called the absolute spectral
gap~\cite{Levin2008}) is defined as $\gamma_* = 1 - \lambda_*$. It corresponds
to the minimum
distance from the unit circle in the complex plane for all eigenvalues
different from $1$. If the transition matrix is aperiodic and irreducible,
then the gap is larger than zero. If the transition matrix is reversible, then
the gap satisfies $\gamma_* = 1-\lambda_2$, where $\lambda_2$ is the
second-largest eigenvalue.
In our example of \eq{equ:SpectrumCollapsed}, the gap is realized by
$\lambda(h=1)$,
\begin{equation}
\gamma_*=
1 - \lambda(h=1) = p\glc 1 -  \cosb{\frac{\pi}{N}} \grc \sim  p \frac{\pi^2
}{2 N^2 },
\label{equ:GapUniformPath}
\end{equation}
which corresponds to the well-known $\sim N^2 $ relaxation time (the inverse
gap) for diffusive motion on an interval of length $N$ (see
\app{app:ExampleTransitionPathFlat} for the explicit $N\times N$ transition
matrix corresponding to \eq{equ:TransitionMatrixCollapsed}, the derivation of
\eq{equ:SpectrumCollapsed} and further details).

In the second step of our three-step path towards non-reversibility,
we introduce loops to the graph.
This is because all Markov chains on trees (graphs without loops) are
reversible. To do so, we \quot{lift} the path graph (that is, duplicate it) into
a  \quot{$-1$} copy and a \quot{$+1$} copy, with
transitions between copies. This results in the $2N$-site \quot{lifted} path
graph (here shown for $N=4$) with the transition probabilities
\begin{equation}
 \begin{tikzcd}[column sep = 1.2cm, row sep = 0.6cm]
 \circled{$1+$} \arrow{r}{\half p}
 \arrow[loop above]{}{\underbrace{\text{\small $1 - p - \epsilon$}}}
\arrow[xshift=-1.0ex,swap]{d}{\epsilon + \half p }
 &
\circled{$2+$} \arrow{r}{\half p}
\arrow[yshift=-1.0ex]{l}{\half p}
 \arrow[loop above]{}{\underbrace{\text{\small $1 - p - \epsilon$}}}
\arrow[xshift=-1.0ex,swap]{d}{\epsilon }
&
\circled{$3+$}
\arrow[yshift=-1.0ex]{l}{\half p}
 \arrow[loop above]{}{\underbrace{\text{\small $1 - p - \epsilon$}}}
\arrow{r}{ \half p}
\arrow[xshift=-1.0ex,swap]{d}{\epsilon } &
\circled{$4+$} \arrow[yshift=-1.0ex]{l}{\half p}
 \arrow[loop above]{}{\underbrace{\text{\small $1 - p - \epsilon$}}}
\arrow[xshift=-1.0ex,swap]{d}{\epsilon + \half p }
\\
\circled{$1-$}\arrow[yshift=1.0ex]{r}{\half p}
 \arrow[loop below]{}{\overbrace{\text{\small $1 - p - \epsilon$}}}
\arrow[xshift=1.0ex,swap]{u}{\epsilon + \half p }
&
\circled{$2-$} \arrow[yshift=1.0ex]{r}{\half p}
\arrow{l}{\half p}
 \arrow[loop below]{}{\overbrace{\text{\small $1 - p - \epsilon$}}}
\arrow[xshift=1.0ex,swap]{u}{\epsilon } &
\circled{$3-$} \arrow[yshift=1.0ex]{r}{\half p}
 \arrow[loop below]{}{\overbrace{\text{\small $1 - p - \epsilon$}}}
\arrow{l}{\half p}
\arrow[xshift=1.0ex,swap]{u}{\epsilon }
&
\circled{$4-$} \arrow{l}{\half p}
 \arrow[loop below]{}{\overbrace{\text{\small $1 - p - \epsilon$}}}
\arrow[xshift=1.0ex,swap]{u}{\epsilon + \half p }
\end{tikzcd}
\label{equ:LiftedTransitionReversible}
\end{equation}
(see \app{app:ExampleTransitionPathLiftedFlat} for the associated $2N\times 2N$
transition matrix and for further details). For $1-p-\epsilon >0$,
\eq{equ:LiftedTransitionReversible}
describes an irreducible, aperiodic and reversible transition matrix, with a
flat steady state. As before, the spectrum can be computed using the Fourier
basis and is given by:
 \begin{align}
 \lambda^\flatshape(h=0)  &  =  1,
 \label{equ:Value0Fan} \\
 \lambda^\flatshape(h=N)  &  =  1 - 2 (p + \epsilon),
 \label{equ:ValueNFan}
 \end{align}
as well as by $(N-1)$ pairs of eigenvalues, for $h \in \SET{1 \TO N-1}$,
\begin{equation}
 \lambda_{\pm}^\flatshape(h)    = (1-p-\epsilon) + p \cosb{\pi\frac{h}{N}} \pm
 \epsilon.
\label{equ:LiftedValueReversible}
\end{equation}
The \quot{$+$} sequence of eigenvalues equals that on the path graph, namely the
fan starting at $1$ for $p=0$. This fan defines the gap of the transition
matrix, which is unchanged with respect to the path graph. For fixed $\epsilon$,
a second fan, identical to the first but displaced in $y$ by $-2\epsilon$
complements the spectrum (see \subfig{fig:Flat_A_B_C}{b}). The eigenvalues
$\lambda^+(h)$ and $\lambda^-(h)$ all differ by the same distance $2 \epsilon$,
and we need only respect the condition $1-p-\epsilon > 0$.

In the third and final step of our procedure, we
tune the reversible transition matrix of \eq{equ:LiftedTransitionReversible}
into non-reversibility without changing the steady
state (the NESS is thus by construction unchanged with respect to the
reversible case of \eq{equ:LiftedTransitionReversible}). To do so, we bias the
rates $p$
oppositely in clockwise  ($+\delta$) and counterclockwise ($-\delta$)
direction
around the graph:
\begin{equation}
 \begin{tikzcd}[column sep = 1.2cm, row sep = 0.8cm]
 \circled{$1+$} \arrow{r}{\half p + \delta}
 \arrow[loop above]{}{\underbrace{\text{\small $1 - p - \epsilon$}}}
\arrow[xshift=-0.5ex,swap]{d}{\half p + \epsilon - \delta}
 &
\circled{$2+$} \arrow{r}{\half p + \delta}
\arrow[yshift=-1.0ex]{l}{\half p - \delta}
 \arrow[loop above]{}{\underbrace{\text{\small $1 - p - \epsilon$}}}
\arrow[xshift=-0.5ex,swap]{d}{\epsilon }
&
\circled{$3+$}
\arrow[yshift=-1.0ex]{l}{\half p - \delta}
 \arrow[loop above]{}{\underbrace{\text{\small $1 - p - \epsilon$}}}
\arrow{r}{ \half p + \delta}
\arrow[xshift=-0.5ex,swap]{d}{\epsilon } &
\circled{$4+$} \arrow[yshift=-1.0ex]{l}{\half p - \delta}
 \arrow[loop above]{}{\underbrace{\text{\small $1 - p - \epsilon$}}}
\arrow[xshift=-0.5ex,swap]{d}{\half p + \epsilon + \delta}
\\
\circled{$1-$}\arrow[yshift=1.0ex]{r}{\half p - \delta}
 \arrow[loop below]{}{\overbrace{\text{\small $1 - p - \epsilon$}}}
\arrow[xshift=0.5ex,swap]{u}{\half p + \epsilon + \delta}
&
\circled{$2-$} \arrow[yshift=1.0ex]{r}{\half p - \delta}
\arrow{l}{\half p + \delta}
 \arrow[loop below]{}{\overbrace{\text{\small $1 - p - \epsilon$}}}
\arrow[xshift=0.5ex,swap]{u}{\epsilon}
&
\circled{$3-$} \arrow[yshift=1.0ex]{r}{\half p - \delta}
 \arrow[loop below]{}{\overbrace{\text{\small $1 - p - \epsilon$}}}
\arrow{l}{\half p + \delta}
\arrow[xshift=0.5ex,swap]{u}{\epsilon }
&
\circled{$4-$} \arrow{l}{\half p + \delta}
 \arrow[loop below]{}{\overbrace{\text{\small $1 - p - \epsilon$}}}
\arrow[xshift=0.5ex,swap]{u}{\half p + \epsilon - \delta}
\end{tikzcd}.
\label{equ:LiftedTransitionNonReversible}
\end{equation}
With the non-reversibility parameter $\delta$,
\eq{equ:LiftedTransitionNonReversible} represents a non-reversible
transition matrix $P$ that is doubly stochastic (any row and any column
sums to one). The steady state is thus again flat. For $\delta
\neq 0$ and
$\epsilon$ small, the diffusive behavior is broken, and there is finite
flow
in clockwise direction for $\delta >0$, and in counterclockwise direction
for
$\delta < 0$. The parameter $\epsilon$ hybridizes the two fans of
\subfig{fig:Flat_A_B_C}{b}.

\begin{figure}[htb]
	\centering
\includegraphics[width=9cm]{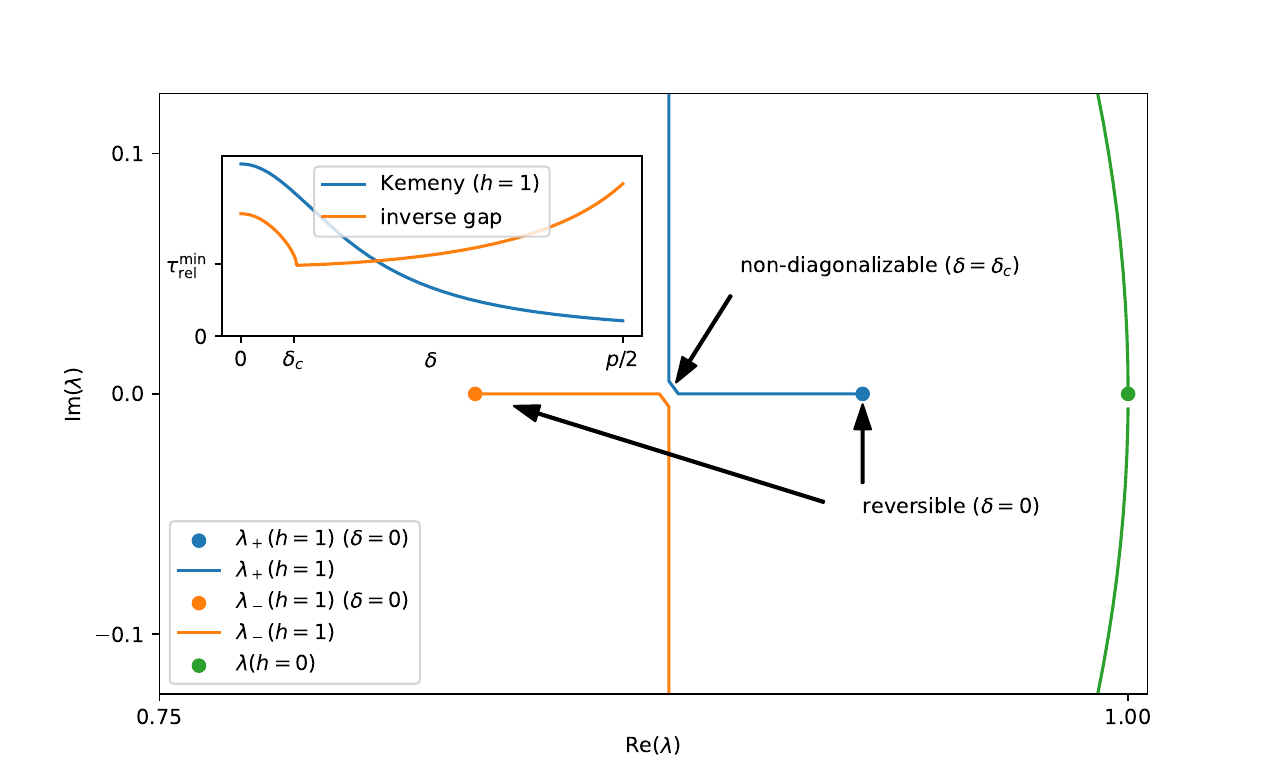}
  \caption{Eigenvalues $\lambda_\pm(h=1)$
  as a function of the non-reversibility $\delta$ (flat steady state, $N=8$,
with $\epsilon=0.05$ and $p=0.9$), as they approach each other on the
real axis,
then separate vertically along the imaginary direction.
The eigenvalue $\lambda_+(h=0) $ and its associated left eigenvector are
independent of $\delta$.
Inset:
The inverse of the gap (the minimum distance to the unit circle, see
\eq{equ:MaxAbs}), thus the
relaxation time $\taurel$, is minimal when the two eigenvalues meet at
$\delta_c$ and the transition matrix is non-diagonalizable. The \Kemeny
$\taukemeny(h=1) =
1/\glc 1-\lambda_+(h=1) \grc + 1/\glc 1-\lambda_-(h=1) \grc$ will be introduced
in \sect{sec:GM_Kemeny}. }
\label{fig:ReversibleNonReversible}
\end{figure}

The transition matrix satisfies, for a flat steady state
$\pi_{(i, \sigma)} = \const$ on all $2N$ sites, the global-balance condition
$\pi_{(i, \sigma)} = \sum_{(j, \sigma')} \pi_{(j, \sigma')} P_{(j,
\sigma'),(i,\sigma)}\quad  \forall (i,\sigma)\in \Omega$ that we will discuss
extensively throughout the present paper. The two lifting conditions are
satisfied~\cite{ChenLovaszPak1999}: First, the sum of the weights
of the lifted site agrees with the weight of the original
site ($\sum_{\sigma }\pi_{i \sigma} = \pi_{i} =
\frac{1}{N}$) and, second, the sum of the lifted flows agrees with
the original flows ($\sum_{\sigma, \sigma'} \pi_{i\sigma} P_{i\sigma, j \sigma'}
= \pi_{i} P_{ij}$).
The spectrum of the transition matrix can again be computed using the Fourier
basis. It consists of two isolated eigenvalues,
 \begin{align}
 \lambda^\flatshape(h=0)  &  =  1,
 \label{equ:Value0} \\
 \lambda^\flatshape(h=N)  &  =  1 - 2 (p + \epsilon),
 \label{equ:ValueN}
 \end{align}
as well as of $N-1$ pairs of eigenvalues, for $h \in \SET{1 \TO N-1}$,
\begin{equation}
 \lambda_{\pm}^\flatshape(h)    = (1-p-\epsilon) + p \cosb{\pi\frac{h}{N}} \pm
\sqrt{
\epsilon^2 - 4 \delta^2 \sinb[2]{\pi \frac{ h}{N}}}
\label{equ:LiftedValue_h}
\end{equation}
(see again \app{app:ExampleTransitionPathLiftedFlat} for the explicit $2N
\times 2N$ transition matrix
corresponding to \eq{equ:LiftedTransitionNonReversible} and the derivation of
\eqfromto{equ:Value0}{equ:LiftedValue_h}).
These eigenvalues can now be complex.
Any two eigenvalues $\lambda_+(h)$ and $\lambda_-(h)$, which at $\delta=0$
differ by $2 \epsilon$, now approach each other as a function of $\delta^2$,
with a \quot{speed} which is smallest for $h=1$ and $h=N-1$ and maximal for
$h=N/2$, then meet in the middle of the interval, from where they move
vertically up and down on the imaginary axis (see
\fig{fig:ReversibleNonReversible}). Clearly, the gap---the distance to the unit
circle---is largest when the two eigenvalues coincide.
The transition matrix is then non-diagonalizable.

For real-valued eigenvalues, the gap $\Delta$ is given by
\begin{equation}
 \Delta = p \glb 1 - \cosa{\frac{\pi}{N}} \grb + \epsilon
\glb 1-\sqrt{1 - \frac{4 \delta^2}{\epsilon^2}  \sina[2]{\frac{\pi}{N}}} \grb.
\label{equ:GapFlatLifted1}
\end{equation}
For fixed $N, p$, and $\epsilon$,
the gap (the distance to the unit circle) is maximum when the
square root in \eq{equ:GapFlatLifted1} vanishes, for
\begin{equation}
\delta_c = \frac{\epsilon}{2\sina{\frac{\pi}{N}}}.
\end{equation}
At $\delta_c$, the two eigenvalues $\lambda_\pm(h=1)$ meet (see
\fig{fig:ReversibleNonReversible}). At this point, the transition matrix
is non-diagonalizable. For $\delta > \delta_c$, the two eigenvalues
move vertically in the complex plane.
With $\epsilon = \epsilontilde / N$, this gives for large $N$:
\begin{equation}
\Delta
\simeq p\frac{\pi^2}{2 N^2} + \frac{\epsilontilde}{N}
\underbrace{\glb 1-\sqrt{1 - \frac{4 \delta^2 \pi^2}{\epsilontilde^2}
} \grb}_{\text{$\neq 0$ if $\delta \neq 0$}}.
\label{equ:GapFlatLifted}
\end{equation}

As is evident from a graphical analysis, the gap of the spectrum, the minimum
distance of the eigenvalues from the unit circle, increases with $\delta$ away
from the reversible case $\delta = 0$, and until $\delta_c$, to then decrease
with $\delta > \delta_c$ (see the inset of \fig{fig:ReversibleNonReversible},
which shows the inverse gap). The minimal relaxation time (inverse gap) is
thus reached where  the dominant eigenvalues meet and the transition matrix is
non-diagonalizable.

The Markov chain with the flat steady state on the
lifted path graph~\cite{Diaconis2000}
contains, in a nutshell, the ingredient that renders non-reversibility
interesting, namely the scaling in $\sim N$, rather than in $\sim N^2$, of the
relaxation time, that comes from the hybridization of the eigenvalues
$\lambda_+$ and $\lambda_-$.
It also illustrates that the spectrum of a non-reversible transition matrix may
be (but does not have to be) complex-valued, and that a non-reversible
transition matrix may be non-diagonalizable.

In the remainder of the present paper, we consider transition matrices on the
path
graphs and lifted path graphs of
\eqfromto{equ:TransitionMatrixCollapsed}{equ:LiftedTransitionNonReversible}, but
for steady states that are not flat. The parameter $\delta$ again drives the
reversible Markov chain into non-reversibility
without changing the steady state. The spectrum of the transition
matrix can then not in general be obtained analytically for large $N$. To
complement our numerical calculations, we consider a matrix-valued  Green's
function, the \quot{\GM}, which in some cases (but not for small $\delta$)
can be computed even when the
spectrum is unknown. Its trace provides the \Kemeny~\cite{doyle2009kemeny}, a
function of all the eigenvalues. We study the relation of the \Kemeny
with the
mixing and hitting times, as well as the inverse gap of the
transition-matrix spectrum.

The \GM also allows one to compute the \MFPT, and we investigate its connection
with relaxation and mixing times. In \sect{sec:BasicFacts}, we provide a number
of definitions for Markov chains omitted for brevity in this introduction and
discuss  the \GM approach for the aforementioned flat steady
state. In \sect{sec:GMKemenyFlat} we compute the Green’s
matrix and the \Kemeny  on the path graph and the lifted path graph. In
\sect{sec:MetropolisPathLiftedPath}, we generalize the transition probabilities
in \eqfromto{equ:TransitionMatrixCollapsed}{equ:LiftedTransitionNonReversible}
to contain an additional Metropolis filter, that accepts or
rejects a proposed move,  and that imposes
non-flat
steady states. Through the \GUM approach we
reach important information on the relaxation times and the \MFPT that we
confirm through \ExactDiag. Our conclusions are in \sect{sec:Conclusions}. A
series of appendices accompanies the present paper. Computer programs (in Python
and in Mathematica) are made available in an open-source repository (see
\app{app:ComputerPrograms}). They follow all the explicit calculations of the
present paper in concrete examples, and also illustrate concepts and notations.

\section{Markov chains: from the transition matrix $P$ to the  \GM $G$}
\label{sec:BasicFacts}

In this section, we discuss the transition matrix of a finite Markov chain
and its characteristic properties and time scale~\cite{Levin2008}, and then
introduce the \GM and the \Kemeny, which we compute for specific models
`in later sections. In the present paper, Markov chains evolve in discrete time
$t=0,1,2,\dots$. The continuous-time formalism has also been
developed~\cite{Mazzolo2023}.

\subsection{Transition matrix, irreducibility, aperiodicity, convergence times}

For a finite sample space $\Omega$ with $N= |\Omega|$,
the transition matrix is an $N \times N$ square matrix whose element $P_{ij}$
gives the conditional probability for
the Markov chain to move from site $i$ to site $j$ in one step, thus
defining a Markov-chain Monte Carlo algorithm (we refer to the elements of the
sample space as \quot{sites}).
If, at time $t$, the Markov chain is at
site $j$ with probability $\pit{t}_j$ for all $j$, then at time $t+1$ it is
at site $i$ with probability $\pit{t+1}_i$:
\begin{equation}
\underbrace{\BRA{\pit{t+1}}
= \BRA{\pit{t}}  P
}_{\text{Dirac notation}}
 \DiracCoord
 \underbrace{
 \pit{t+1}_i = \sum_{j \in \Omega} \pit{t}_j P_{ji}
 \quad \forall i \in \Omega}_{\text{component notation}}
 \label{equ:TimeEvolution}
\end{equation}
(see \app{app:Dirac} for a guide to Dirac notations).
Global balance,
\begin{equation}
 \BRA{\pi} = \BRA{\pi}  P \DiracCoord \pi_i = \sum\nolimits_{j \in \Omega}
\pi_j P_{ji}
 \quad \forall i \in \Omega,
\label{equ:GlobalBalance}
\end{equation}
follows as a condition on the transition matrix from \eq{equ:TimeEvolution} if
we impose a steady state $\pi$.
This imposed steady state $\BRA{\pi} \DiracCoord
\SET{\pi_1 \TO
\pi_N}$ is a real-valued left eigenvector with eigenvalue $\lambda_1=1$ of $P$.
The eigenvalue $\lambda_1=1$ is non-degenerate, and the steady state
$\BRA{\pi}$ is unique if $P$ is irreducible, that is,
if any site $i$ can be reached from any
$j$ in a finite time. The eigenvalue $\lambda_1$ is the unique eigenvalue of
unit norm and $\BRA{\pi}$ is reached from any starting distribution
$\BRA{\pit{t=0}}$ if the transition matrix is irreducible and aperiodic.

Transition matrices are stochastic: they satisfy
\begin{equation}
 \KET{1} = P\KET{1} \DiracCoord \sum\nolimits_{j \in \Omega}   P_{ij} =1
 \quad \forall i \in \Omega,
\label{equ:Stochasticity}
\end{equation}
simply because the Markov chain, between times $t$ and $t+1$, has to move to
another site or else stay at $i$. The \eq{equ:Stochasticity} means
that $\KET{1} \DiracCoord \SET{ 1 \TO 1}^{T}$ is a right eigenvector of $P$
with eigenvalue $\lambda_1 = 1$.

An irreducible, aperiodic Markov chain converges from any initial
site $i$ towards the
steady state.  This means that the matrix $\Delta(t)$,
\begin{equation}
\Delta(t) \equiv P^t - \KET{1}\BRA{\pi} \DiracCoord
\Delta_{ij}(t) = (P^t)_{ij} - \pi_j \quad \forall i,j \in \Omega,
 \label{equ:defdelta}
\end{equation}
approaches zero ($\Delta(t) \to \ZERO$) for large times $t$. As
$\KET{1}\BRA{\pi}$ is a positive matrix (see \app{app:DiagonalizableMatrix}
for an illustration), the matrix $P^t$, for large enough
$t$, is also a positive matrix, in
other words:
\begin{equation}
\text{for $t > t_0$:}\quad
\BRA{i} P^t \KET{j} > 0  \DiracCoord (P^t)_{ij} >0
\quad \forall i,j \in \Omega.
\label{equ:Propagatorij}
\end{equation}
We refer to $\langle i \vert P^t \vert j \rangle$ as the propagator from
$i$ to $j$ in $t$ steps.

If the matrix $P$ is diagonalizable, which is the case, among others, when it
has distinct eigenvalues, it has a spectral decomposition
\begin{equation}
 P  = \sum_{m=1}^N \lambda_m \RMLM
 = \ONEPI + \sum_{m\ne 1} \lambda_m \RMLM
\label{equ:chainspectral}
\end{equation}
with right eigenvectors $\KET{\R{m}}$ that satisfy
\begin{align}
 P \KET{\R{m}} = \lambda_m \KET{\R{m}} &\DiracCoord
 \sum_{j \in \Omega }P_{ij} \R{m}_j = \lambda_m \R{m}_i \quad \forall i \in
\Omega,
\label{equ:DefEigenVectorRight}
 \intertext{and left eigenvectors $\BRA{\L{m}}$ that
satisfy}
 \BRA{\L{m}}  P =\lambda_m   \BRA{\L{m}} &\DiracCoord
 \sum_{i \in \Omega } \glc \L{m}_i\grc^*  P_{ij}  = \lambda_m \glc \L{m}_j
\grc^* \quad \forall
j \in \Omega.
\label{equ:DefEigenVectorLeft}
\end{align}
The subtlety about complex conjugates arises because, by convention, a vector
$l$ is a left eigenvector of a matrix $P$ with eigenvalue $\lambda$ if $\sum_i
l_i^* P_{ij}  = \lambda l_j^*$, where \quot{$\ ^* \
$} denotes complex conjugation.

The left and right eigenvectors with eigenvalue $\lambda_1 = 1$ are special,
and we have
\begin{equation}
\BRA{\L{1}}  \equiv
\BRA{\pi} \quad \KET{\R{1}} \equiv \KET{1}.
\end{equation}
Left and right eigenvectors satisfying the orthonormalization and closure
relations
\begin{align}
 \BRAN{\L{m}} \KET{\R{m'}} &= \delta_{m,m'} \DiracCoord
 \sum_{i \in \Omega} \glc \L{m}_i \grc^* \R{m'}_i = \delta_{m, m'}
\nonumber  \\
 \ONE  &=  \sum_m \KET{ \R{m}} \BRA{\L{m}},
\label{equ:fermeturej}
\end{align}
that are further illustrated in the example programs of \app{app:MathDetails}.

\subsection{\GM, \MFPT,  \Kemeny}
\label{sec:GM_Kemeny}

As mentioned, the matrix $\Delta(t)$ of differences between $P^t$ and the
steady state
approaches zero for $t\to \infty$. The \emph{\GM}, a generalized Green's
function, sums up these differences for all times,
\begin{equation}
G = \sum_{t=0}^\infty \Delta(t)
 =  \sum_{t=0}^{\infty} \glb P^t - \ONEPI \ \grb
 =  \sum_{m \neq 1} \frac{1}{1 - \lambda_m} \RMLM,
\label{equ:GreenDef}
\end{equation}
where, on the right-hand-side transformation, we again
suppose that $P$ is diagonalizable (non-diagonalizable
$P$ correspond to special points in parameter space and add nothing new in our
context).

The \GM satisfies several linear relations. First, it commutes with the
transition matrix, $GP=PG$, and its
orthogonality relations are
\begin{align}
\ZERO =  \BRA{\pi}  G & \DiracCoord 0 =  \sum_{i \in \Omega} \pi_i G_{ij}
\quad  \forall j \in \Omega,
 \label{equ:LeftGorthog}  \\
\ZERO =    G \KET{1} &  \DiracCoord 0 = \sum_{j \in \Omega}  G_{ij} \quad
\forall i \in \Omega
\label{equ:Gorthog}
\end{align}
(see \app{app:Dirac}).
Its trace, known as the \Kemeny
$\taukemeny$~\cite{doyle2009kemeny,Hunter2014,Bini2018,Pinsky2019},
is expressed through the spectrum of the transition matrix:
\begin{equation}
\taukemeny: = \Tr\ G = \sum_{m \neq 1}\frac{1 }{1 - \lambda_m}.
\label{equ:TraceGreen}
\end{equation}
It follows from \eq{equ:GreenDef}, using $P ^0 = \ONE$,
that
\begin{equation}
\ONE - \ONEPI = G - PG \DiracCoord \delta_{ij} - \pi_j = G_{ij} - \sum_{k \in
\Omega}
P_{ik} G_{kj} \quad \forall i, j \in \Omega.
\label{equ:GminusPG}
\end{equation}
Here, the left-hand sides are explicitly known through the steady state
while the right-hand sides connect the \GM  and the transition
matrix. It follows that
\begin{equation}
\delta_{ij} - \pi_j   = \underbrace{\glb \sum_{k\in \Omega} P_{ik} \grb}_{=1}
G_{ij} - \sum_{k
\in \Omega} P_{ik}
G_{kj}
=  \sum_{k \ne i} P_{ik} (G_{ij} -  G_{kj} ) \quad \forall i,j \in \Omega.
\label{equ:DifferenceOfG}
\end{equation}
The transition matrices considered in the present paper are
sparse,
and \eq{equ:DifferenceOfG} turns into a recursion for the differences
of the \GM that can often be rendered explicit.
In addition to the recursion
for differences of the \GM,
the left-sided orthogonality relation of \eq{equ:LeftGorthog} expresses
individual \GUM elements through these differences:
\begin{equation}
0 =   \sum_{i \in \Omega} \pi_i G_{ij}  = \pi_j G_{jj} +  \sum_{i \ne j}  \pi_i
G_{ij}
= \glb 1- \sum_{i \ne j}  \pi_i\grb G_{jj} +  \sum_{i \ne j}  \pi_i
G_{ij} \quad  \forall j \in \Omega.
\label{equ:Gorthogj}
\end{equation}
This implies that
\begin{equation}
   G_{jj} = - \sum_{i \ne j}  \pi_i (G_{ij}-G_{jj}) \quad \forall j \in \Omega.
\label{equ:GorthogjjDiff}
\end{equation}

We next consider the \MFPT $\tau_{ij}$ from site $i$ to site $j$:
 \begin{equation}
\tau_{ij} \equiv \sum_{t_1=0}^{+\infty}  t_1 F_{ij}(t_1) \quad \forall i,j \in
\Omega
\label{equ:mfpt}
\end{equation}
with the hitting probability $F_{ij}(t_1)$ of visiting $j$ for the first time at
time $t_1$ when starting at $i$ at time $0$. We will express any element
$G_{ij}$ of the \GM in terms of $\tau_{ij}$ and $\pi_{i}$ and thus connect
the \GM to the \MFPT, that is, to the physical behavior of the
system.

For an irreducible transition matrix which eventually comes to any site $j$
from any site $i$, the hitting probabilities satisfy the normalization
and $t=0$
initial conditions
\begin{equation}
\sum_{t_1=0}^{+\infty} F_{ij}(t_1)  =1; \quad
F_{ii}(t_1) =\delta_{t_1,0}; \quad \tau_{ii}  = 0.
\label{equ:firstpassagedistri}
\end{equation}
The propagator $\BRA{ i} P^t \KET{j}$ (see \eq{equ:Propagatorij}) writes as the
hitting
probability $F_{ij}(t_1)$ to return to $j$ for the first time after  $t_1$
steps and then
to return from $j$ to $j$
in the remaining time $(t-t_1)$:
\begin{equation}
\BRA{i} P^t \KET{j} = \sum_{t_1=0}^t F_{ij}(t_1)\BRA{j}
P^{t-t_1} \KET{j} \quad \forall i,j \in \Omega.
\label{equ:propagatorandfirst}
\end{equation}
The propagator $\BRA{i} P^t \KET{j}$ on the left-hand side involves
$\pi_j $ and $\Delta_{ij}(t)$, while the propagator on the right-hand side
again involves $\pi_j$ and $\Delta_{jj}(t-t_1)$ (see \eq{equ:defdelta}).
Summed over all times $t$, the terms in $\Delta$ express in terms of the \GM,
\begin{equation}
 G_{ij} = - \tau_{ij} \pi_j + G_{jj} \quad \forall i,j \in \Omega,
\label{equ:GFirstPassage}
\end{equation}
and relate the first-passage time $\tau_{ij}$ to a difference between two
Green-matrix elements (see \app{app:GM_MFPT} for the derivation of
\eq{equ:GFirstPassage}).
Summing over $j$, the left-hand side vanishes (see \eq{equ:Gorthog}), while the
right-hand side, which then cannot depend on $i$, again yields the trace of
the \GM, in other words, the \Kemeny,
\begin{equation}
\taukemeny =  \sum_{j\in \Omega} \tau_{ij} \pi_j = \Tr\ G = \sum_{m \neq
1}\frac{1 }{1 - \lambda_m}\quad \forall i \in \Omega,
\label{equ:KemenySpectrumMFPT}
\end{equation}
where, on the right, we suppose a diagonalizable transition matrix.
The above formula produces identical output for all $i \in
\Omega$, a fact also known as the \quot{random target lemma}~\cite[Lemma
10.1]{Levin2008}.
It relates the trace of  \GM to the matrix of the
\MFPTs $\tau_{ij}$.

The left-sided orthogonality condition allows one to write the
diagonal
\GUM element $G_{jj} $ in terms of the
MFPT $\tau_{ij}$ and the steady state $\pi$ as
\begin{equation}
  \underbrace{ G_{jj} = -   \sum_{k \in \Omega}  \pi_k
\overbrace{(G_{kj}
-G_{jj})}^{\text{\eq{equ:GFirstPassage}}}}_{\text{\eq{equ:GorthogjjDiff}}}
= \hspace{-0.3cm}
\underbrace{\glb \sum_{k \in \Omega }  \pi_k
\tau_{kj} \grb}_{\tau^{\text{eq.}}_j\text{: equilibrium access time of $j$}}
\hspace{-0.7cm}
\pi_j \quad \forall j \in \Omega
\label{equ:Gorthogjtau}
\end{equation}
that involves the equilibrium access time of $j$, in other words the
\MFPT $\tau_{kj}$ towards $j$ from a site $k$ taken from
equilibrium. Then, \eq{equ:GFirstPassage} can be rewritten as
\begin{equation}
G_{ij} = \glb \sum_{k \in \Omega }  \pi_k \tau_{kj} -  \tau_{ij}\grb \pi_j
\quad \forall i,j \in \Omega.
\label{equ:greenandfirsttauonly}
\end{equation}
This equation expresses any element of the \GM in terms of physical quantities.
It reduces to \eq{equ:Gorthogjtau} for $i=j$, because $\tau_{jj}= 0$.

\subsection{Relation between \Kemeny, relaxation time, set time, and mixing
time}
\label{sec:ConductancesRelaxationsSetTimes}

The \Kemeny, from its representation of \eq{equ:KemenySpectrumMFPT} in terms of
all eigenvalues,
is larger than the inverse of the gap if the latter is realized with a
positive real eigenvalue ($| \lambda_2|= \lambda_2$).
Indeed, in this case, we have
\begin{equation}
   \taukemeny = \frac1{1 - |\lambda_2|} + \sum_{k > 2} \frac1{1- \lambda_k} \geq
\frac1{1 - |\lambda_2|}.
\end{equation}
If  $\lambda_2 $ has a finite imaginary component, there is no definite ordering
of the \Kemeny and the relaxation time, as is evident for the flat steady state
on the lifted path graph (see inset of \fig{fig:ReversibleNonReversible}). We
will witness a number of irreducible and aperiodic transition matrices where the
\Kemeny has better scaling than the relaxation times. For an irreducible
periodic transition matrix, the \Kemeny is in general finite while the inverse
gap is infinite.

As the \Kemeny of \eq{equ:KemenySpectrumMFPT} is independent of the
site $i$, we may write it as
\begin{equation}
 \taukemeny = \sum_{i,j \in \Omega} \pi_i \tau_{ij} \pi_j,
\label{equ:KemenyAsDoubleExpectation}
\end{equation}
(this is \REF[eq. (10.4)]{Levin2008})
in other words as the expected time to hit $j$  from $i$ for a pair of sites
independently sampled from the steady state. This time generically
scales
with $|\Omega|$ (that is, with the cover time), as is evident from the spectral
representation of
\eq{equ:KemenySpectrumMFPT}, which is composed of $|\Omega|$ terms.
During such a long time,
all elements of the sample space can be scanned and, in physicists'
terminology,
the exact calculation of the free energy and the partition function becomes
possible.
In a generic Monte Carlo context however (but not on the path graph or the
lifted path graph), the approach towards equilibrium takes
place on mixing-time and relaxation-time
scales that are much shorter than times that correspond to a complete
scan.
The mixing times (in rough physics terms) specifies the time it
takes to reach  \emph{an arbitrary one}, $j$,  among the many equilibrium
sites
starting from a worst-case site $i$,
but not the time it takes to reach a specific
$j$ (as in \eq{equ:KemenyAsDoubleExpectation}).
The approach of equilibrium
is also encoded in the \GM (but not in its trace).
The diagonal element $G_{jj}$ in
\eq{equ:Gorthogjtau} contains the access time $\tau_{kj}$ of site $j$ from a
site $k$ drawn from equilibrium. In \REF{ChenLovaszPak1999}, the access time is
defined,
not for a single site $j$ but  for
a set $S$, again starting from a site $i$ sampled from the steady state:
\begin{equation}
\tau_{S}^{\text{eq.} } = \text{equilibrium access time of set $S$} =
\sum_{i \in \Omega} \pi_i \tau_{i S} \quad \forall S \subset \Omega.
\end{equation}
Furthermore, the set time of a Markov chain is defined in
\REF{ChenLovaszPak1999} as
\begin{equation}
\tau^{\text{set}}  = \max_{S \subset \Omega}
\glb \tau_{S}^{\text{eq.} } \pi_S \grb.
\end{equation}
It thus follows that the maximum diagonal element of the \GM,
\begin{equation}
 \max_{j} G_{jj} = \max_j \glb \tau_j^{\text{eq.}} \pi_j\grb,
\end{equation}
is a set time of the Markov chain, restricted to single-element sets. The set
time, at least in its general definition, is intricately related to the  mixing
time~\cite{ChenLovaszPak1999}, and generically much smaller than the \Kemeny. We
expect the maximum diagonal \GUM element to also be connected to the mixing
time.

\section{Flat steady state on the path graph and the lifted path graph}
\label{sec:GMKemenyFlat}
At the extremal path-graph sites  $i=1$ or $i=N$, the sums on
the
right-hand side of \eq{equ:DifferenceOfG} contain only a single term, resulting
in an explicit formula for some differences of \GUM elements. Inside
the path graph, the sum contains two terms. Starting from one extremity, this
sets up a recursion for differences of \GUM elements, as we will discuss for a
general steady state in \sect{sec:GMKemenyPath}, before specializing
to the flat steady state. On the lifted path graph, the recursion for \GUM
elements can be set up following the same logic if certain transition
probabilities vanish (\sect{sec:GMKemenyLiftedPath}).

In the present section, we introduce the underlying concepts and apply them
to the flat steady state. We will analyze  them for non-constant
steady states in \sect{sec:MetropolisPathLiftedPath}.

\subsection{\GM and \Kemeny, flat steady state, path graph}
\label{sec:GMKemenyPath}
The \GUM approach developed in \sect{sec:GM_Kemeny} yields an explicit
expression for the \Kemeny of
any irreducible transition matrix $P$ on the path graph of
\eq{equ:TransitionMatrixCollapsed}, where the transition probabilities
$\half p$ are generalized into:
\begin{equation}
P_{ij} = P_{i,i+1} \delta_{j,i+1} + P_{i,i-1} \delta_{j,i-1} + (1 - P_{i,i+1} -
P_{i,i-1} ) \delta_{ij},
\label{equ:PathGraphP}
\end{equation}
where $P_{1,0} = P_{N, N+1}=0$. We note that, on the path graph, the steady
state $\pi$ is given explicitly through the transition matrix, as the
detailed-balance condition of \eq{equ:DetailedBalancePath}
yields
$\pi_2$ as a function of $\pi_1$ (known up to normalization), then $\pi_3 $
as a function of $\pi_2$, \etc\
(all weights can be normalized once $\pi_N$ has been computed).

On sites $i=1$ and $i=N$, the transition matrix has only a single finite
off-diagonal term which, from \eq{equ:GminusPG}, initializes a recursion
for differences of \GUM elements, that is continued on the interior sites.
This leads to the following result for the \MFPT $\tau_{mj}$ for
$m<j$:
\begin{equation}
 \tau_{mj} =    \sum_{i=m}^{j-1}\frac{ 1 }{ \pi_i P_{i,i+1} } \sum_{k=1}^i \pi_k
\quad \text{for  $m \in \SET{1 \TO j-1}$}.
\label{equ:tausol}
 \end{equation}
See \app{app:MFPT_Path} for a derivation of \eq{equ:tausol} from \GUM
differences, and
for a corresponding equation for $m > j$.
Using the left of \eq{equ:KemenySpectrumMFPT}, we can then compute
the \Kemeny starting, for instance, from the site $i=1$:
\begin{equation}
\taukemeny  = \tau_{i=1}^* = \sum_{j=1}^N  \tau_{1j} \pi_j
=  \sum_{j=1}^N   \pi_j  \sum_{i=1}^{j-1}\frac{\Pi_i}{ \pi_i P_{i,i+1} }
=  \sum_{i=1}^{N-1}\frac{ \Pi_i }{ \pi_i P_{i,i+1} } \glb 1 -
\Pi_i \grb
\label{equ:KemenyPath}
\end{equation}
(with $\Pi_i = \sum_{k=1}^i \pi_k$). The re-ordered summation on the right of
\eq{equ:KemenyPath} evaluates $\taukemeny$ in  $\sim N$ operations for an
arbitrary
steady state on the path graph.

For the flat steady state on the path graph, we have $\pi_1 = 1/N, P_{i,
i+1} = p/2, \Pi_i = i/N$, so that the \Kemeny is given by
\begin{equation}
\taukemeny =
\frac{2}{p}
\sum_{i=1}^{N-1}
\frac{i (N-i)}{N}
= \frac{1}{p}\frac{N^2 - 1}{3}.
\label{equ:KemenyPathuniform}
\end{equation}
The flat transition-matrix spectrum is given by
\eq{equ:SpectrumCollapsed}, so that the \Kemeny can be computed by the inverse
gaps as
\begin{equation}
\taukemeny = \sum_{m \neq 1} \frac{1}{1-\lambda_m} =
\frac{1}{p}\sum_{h=1}^{N-1} \frac{1}{1 - \cosb{\pi h / N}} =
\frac{1}{p}\frac{N^2 -1}{3},
\label{equ:KemenyPathGraphSpectrum}
\end{equation}
with the eigenvalues from \eq{equ:SpectrumCollapsed}. The \Kemeny of
\eq{equ:KemenyPathuniform} proves the
non-trivial sum in \eq{equ:KemenyPathGraphSpectrum} (see \app{sec:SumEvaluation}
for an independent derivation). This result is in agreement with a diffusive
process occurring on the path graph. In
\sect{sec:MetropolisPathLiftedPath}, we
will compute the \Kemeny for non-constant steady states on the path graph.

\subsection{\GM and \Kemeny, flat steady state, lifted path graph}
\label{sec:GMKemenyLiftedPath}

On the lifted path graph of \eq{equ:LiftedTransitionNonReversible}, we
can derive recursions for the \GM, and then compute the \Kemeny
for the special case $\delta = p/2$, where the lifted loop can
only be traversed clockwise. In the present section, we
discuss this for the flat steady state before generalizing the
approach in \sect{sec:MetropolisPathLiftedPath} to non-constant
steady states.

For concreteness, we show again the transition matrix on the lifted path graph,
for the maximal non-reversibility $\delta= p/2$:
\begin{equation}
 \begin{tikzcd}[column sep = 1.2cm, row sep = 0.8cm]
 \circled{$1+$} \arrow{r}{p}
 \arrow[loop above]{}{\underbrace{\text{\small $1 - p - \epsilon$}}}
\arrow[xshift=-0.5ex,swap]{d}{\epsilon}
 &
\circled{$2+$} \arrow{r}{p}
 \arrow[loop above]{}{\underbrace{\text{\small $1 - p - \epsilon$}}}
\arrow[xshift=-0.5ex,swap]{d}{\epsilon }
&
\circled{$3+$}
 \arrow[loop above]{}{\underbrace{\text{\small $1 - p - \epsilon$}}}
\arrow{r}{p}
\arrow[xshift=-0.5ex,swap]{d}{\epsilon } &
\color{red}{{\circled{$4+$}}}
 \arrow[loop above]{}{\underbrace{\text{\small $1 - p - \epsilon$}}}
\arrow[xshift=-0.5ex,swap]{d}{\color{red}{p + \epsilon}}
\\
\color{red}{\circled{$1-$}}
 \arrow[loop below]{}{\overbrace{\text{\small $1 - p - \epsilon$}}}
\arrow[xshift=0.5ex,swap]{u}{\color{red}{p + \epsilon}}
&
\circled{$2-$}
\arrow{l}{p}
 \arrow[loop below]{}{\overbrace{\text{\small $1 - p - \epsilon$}}}
\arrow[xshift=0.5ex,swap]{u}{\epsilon}
&
\circled{$3-$}
 \arrow[loop below]{}{\overbrace{\text{\small $1 - p - \epsilon$}}}
\arrow{l}{p}
\arrow[xshift=0.5ex,swap]{u}{\epsilon }
&
\circled{$4-$} \arrow{l}{p}
 \arrow[loop below]{}{\overbrace{\text{\small $1 - p - \epsilon$}}}
\arrow[xshift=0.5ex,swap]{u}{\epsilon}
\end{tikzcd}.
\label{equ:LiftedTransitionNonReversibleSpecial}
\end{equation}

\subsubsection{\Kemeny from \GM, flat steady state, lifted path graph}
In the transition matrix associated with the graph of
\eq{equ:LiftedTransitionNonReversibleSpecial}, the counterclockwise
loop has
disappeared with respect to \eq{equ:LiftedTransitionNonReversible}.
The upper right and lower left
sites (shown in red) have unique non-diagonal outgoing arrows. Just as on the
path graph, this sets off a recursion for the difference of \GUM elements, and
allows us to compute the \MFPT $\tau^{\sigma\sigma'}_{mj}$. For
our purposes, we only require the term for $\sigma=1$ and $m=1$
\begin{align}
\tau_{1k}^{+,\sigma'}  & =
\sum_{m=1}^{k-1} \frac{1}{p + \epsilon}\glb  1 + \frac{2m\epsilon}{p}\grb
+ \frac{\delta^{\sigma', -} }{p+\epsilon} \glc 1 + 2 (N-k) \grc
\quad \text{for  $k > 1$}, \\
\tau_{11}^{+,-}  & = \frac{2N - 1}{p + \epsilon}
\label{equ:tausolLiftedPath}
\end{align}
(see \app{app:MFPT_LiftedPath} for a derivation of
\eq{equ:tausolLiftedPath}
from \GUM differences for the topology of
\eq{equ:LiftedTransitionNonReversibleSpecial}, but with more general transition
rates).
Using the left of \eq{equ:KemenySpectrumMFPT}, we can again compute
the \Kemeny starting, for instance, with the site $i=(1,+)$:
%
%
 \begin{align}
  \taukemeny & =  \sum_{k=1}^N \sum_{\sigma''=\pm}  \tau^{+ \sigma''}_{1k}
\frac{\pi_k}{2}
 = \tau^{+ -}_{11} \frac{\pi_1}{2} + \sum_{k=2}^N \sum_{\sigma''=\pm}  \tau^{+
\sigma''}_{1k} \frac{\pi_k}{2}
 \nonumber \\
&=
\frac{2\epsilon}{p(p+\epsilon)} \sum_{k=1}^{N-1}
\glc 1 - \frac{1}{N}(N-k)^2 \grc + \frac{1}{p } \sum_{k=1}^{N}
\glb \frac{1}{2N} + \frac{2}{N} \sum_{j=k+1}^{N} 1 \grb \\
&=
\frac{\epsilon}{p(p+\epsilon)}
\frac{N^2 -1 }{3} + \frac{1}{p+\epsilon} \glb N - \half \grb
\label{equ:KemenyLiftedPathGreen}
 \end{align}
(see \app{app:MathDetails} for a detailed derivation for arbitrary steady
state).
For $\epsilon \sim 1/N$, the \Kemeny scales as $\taukemeny \sim N$.
It is minimal  for vanishing resampling rate $\epsilon = 0$,
and then still scales as $\sim N$. In that case $\epsilon=0$, the inverse gap
scales as $\sim N^2$ for $p<1$ and is infinite for $p=1$, where $P$
is periodic. This again illustrates that the \Kemeny is more closely related to
a hitting time, as discussed in \sect{sec:ConductancesRelaxationsSetTimes}, than
to the relaxation time (see also \REF{Pinsky2019}).

\subsubsection{\Kemeny from spectrum, flat steady state, lifted path graph}

In \sect{sec:MetropolisPathLiftedPath}, we will study the transition matrix of
\eq{equ:LiftedTransitionNonReversibleSpecial} generalized to non-constant
steady states. For the flat steady state on the lifted path graph, the
\Kemeny of \eq{equ:KemenyLiftedPathGreen} can be tested through the
inverse-gap formula of \eq{equ:TraceGreen}.
Indeed, for vanishing counterclockwise flow $\delta = p/2$,
the paired eigenvalues are
\begin{equation}
 \lambda_{\pm}^\flatshape(h)    = (1-p-\epsilon) + p \cosb{\pi\frac{h}{N}} \pm
\sqrt{
\epsilon^2 - p^2 \sinb[2]{\pi \frac{ h}{N}}}
\label{equ:LiftedValue_hSpecial}
\end{equation}
(here reproduced for clarity from \eq{equ:LiftedValue_h}). They satisfy
\begin{align}
\lambda_+(h) + \lambda_-(h) & = 2 - 2 (p + \epsilon) + 2 p \cosb{\pi\frac{h }
{N}},  \\
\lambda_+(h) \cdot \lambda_-(h) & = 1 - 2 \epsilon + 2 p (- 1 + p + \epsilon)
                    - 2p(-1 + p + \epsilon) \cosb{\pi \frac{h}{N}}.
\end{align}
Any of the pairs of eigenvalues (for $h = 1 \TO N-1$) contributes to the
\Kemeny as follows:
\begin{align}
\taukemeny(h) =
\frac{1}{1 - \lambda_+(h)} +
\frac{1}{1 - \lambda_-(h)}  &=  \frac{2 - \lambda_+(h) - \lambda_-(h)}
   {1
   - \lambda_+(h)
   - \lambda_-(h)
   + \lambda_+(h)\lambda_-(h)
   } \\
   & = \frac{1}{p + \epsilon} + \frac{\epsilon}{(p + \epsilon) p \glc 1 -
\cosb{\pi \frac{h}{N}} \grc }.
\end{align}
Together with the contribution $\taukemeny(N) = \frac{1}{2(p+\epsilon)}$ of the
eigenvalue  $\lambda(h=N) = 1 - 2(p + \epsilon)$ (see \eq{equ:ValueN}), we find
\begin{align}
\taukemeny  =&  \taukemeny (N) + \sum_{h=1}^{N-1} \taukemeny(h)  \\
        =& \frac{1}{2(p + \epsilon)} + \frac{N-1}{p+\epsilon}
+ \frac{\epsilon}{p (p + \epsilon)} \underbrace{\sum_{h=1}^{N-1} \frac{1}{1 -
\cosb{\pi\frac{h}{N}}}}_{\text{see \eq{equ:SumCompute}}}
\label{equ:SumKemenyLifted}\\
=& \frac{1}{p+\epsilon} \glb N - \half + \frac{\epsilon}{p} \frac{N^2
- 1}{3} \grb \quad
\label{equ:KemenyLiftedPathSpectrum}
\end{align}
(see \app{app:ComputerPrograms} for a Mathematica notebook retracing the
derivation of the \Kemeny in this case). For a scaling of $\epsilon \sim 1/N$,
the \Kemeny scales as $\sim N$. This formula agrees with
\eq{equ:KemenyLiftedPathGreen}, which was
obtained with the \GM approach.


\section{General Markov chains on path graphs and lifted path graphs}
\label{sec:MetropolisPathLiftedPath}

\begin{figure}[htb]
	\centering
\includegraphics[width=\columnwidth]{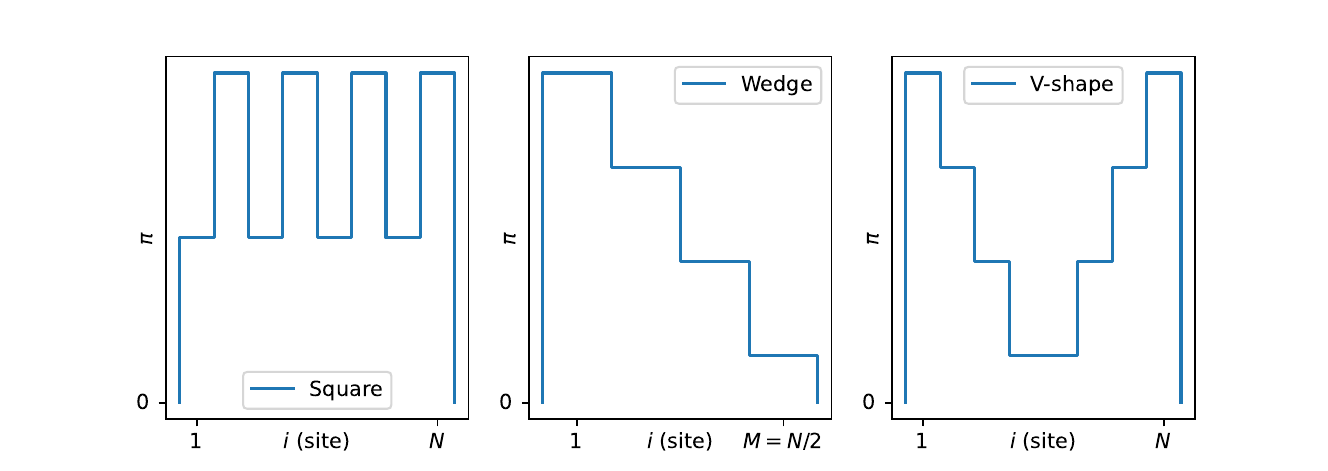}
  \caption{Steady states treated in \sect{sec:MetropolisPathLiftedPath} for
$N=8$.
\subcap{a} Square-wave steady state (see \eq{equ:DHNSquareWave}). On the path
graph, the relaxation time is generically $\taurel \sim N^2$ while, on the
lifted path graph we reach $\taurel \sim N$.
\subcap{b} \WEDGECAP steady state (see \eq{equ:PiWedgeDef}). On the path graph,
generically, we have $\taurel \sim M^2$, and $\taurel \sim M $ on the lifted
path graph. The
\MFPT
to reach site $M$ starting from site $1$ scales as $\sim M^2 \log M$ on the
path graph and as $\sim M^2 $ on the lifted path graph.
\subcap{c} \V steady state (see \eq{equ:PiVDef}). On the path graph,
generically, $\taurel \sim N^2 \log N$ while on the lifted path graph, we have
$\taurel \sim N^2$. These times have an evident interpretation in terms of the
\MFPT, as  equilibrating the \V steady state requires the passage through sites
$M$ and $M+1$, governed by the \MFPT for the \WEDGE steady state.
}
\label{fig:Schema}
\end{figure}

In this section, we generalize the analysis  of the previous sections to
the case of
more general steady states $\SET{\pi_1 \TO \pi_N}$
(\sect{sec:GeneralTransitionMatrices}). We then treat three types of steady
states on the path graph and the lifted path graph, namely the square-wave
(\sect{sec:SquareWave}), as well as the \WEDGE-shape and $V$-shape steady
states (\sect{sec:WedgeVshape}). These cases are illustrated in
\fig{fig:Schema}. For an overview of our findings in this section, see
\fig{fig:Schema}, with the \Kemeny $\taukemeny$ generically (but not always)
scaling as the inverse gap $\taurel$.

\subsection{General transition matrices on the path graph and the lifted
path graph}
\label{sec:GeneralTransitionMatrices}

We now consider Markov chains on the path graph and the lifted path graph
with a non-constant steady state $\pi = \SET{\pi_1 \TO \pi_N}$,
realized with detailed balance or in
non-equilibrium (NESS).
We again propose moves to the left and to the right
with probability $p/2$, but add a Metropolis filter to accept
or reject the move. We again lift the path graph, and introduce the
non-reversibility parameter $\delta$ as in \sect{sec:Introduction}. To
see how a
general reversible Markov chain is tuned into
non-reversibility, we follow the three steps of the procedure  already used in
\sect{sec:Introduction}.

\subsubsection{Metropolis filter on the path graph}
In the first step of the three-step procedure towards non-reversibility, we
again consider the $N$-site path graph and propose to move
to the left and to the right with probability $p/2$, but with a
steady-state-dependent Metropolis filter:
\begin{equation}
 P_{i, i\pm 1}  = \underbrace{\frac{p}{2}}_{\text{proposal}}
\underbrace{\minb{1, \frac{\pi_{i\pm 1}}{\pi_i}}}_{\text{Metropolis filter}}
\quad \forall
i \in \SET{1 \TO N},
\label{equ:MetropolisPath}
\end{equation}
with \quot{phantom} sites  $0$ and $N+1$ with zero weights $\pi_0 = \pi_{N+1} =
0$ at positions  $i=0$ and $i=N+1$, that simplify notations and descriptions of
algorithms.
On the path graph, the above filter implements the reversible
Metropolis algorithm and imposes the steady state $\pi$
\begin{equation}
  \begin{tikzcd}[column sep = 3cm]
\circled{$i-1$
} \arrow[yshift = 0.5 ex]{r}{\half p \minfunc{i}{i-1}}&
 \arrow[loop above]{}{\underbrace{1 - \tfrac12 p
 [ \minfunc{i-1}{i} + \minfunc{i+1}{i}]}}
\circled{ \;\;
$i$
\;\;
 }
\arrow[yshift = 0.5 ex]{r}{\half p \minfunc{i+1}{i}}
\arrow[yshift = - 0.5 ex]{l}{\half p \minfunc{i-1}{i} }
& \circled{$i+1$}  \arrow[yshift = -0.5 ex]{l}
{ \half p \minfunc{i}{i+1}}
\end{tikzcd} .
\label{equ:TransitionMatrixGeneralCollapsed}
\end{equation}
The flows between $i$ and $i + 1 $ balance as do those between $i-1 $ and $i$
and the detailed-balance condition of \eq{equ:DetailedBalancePath} is satisfied
by construction.

\subsubsection{Metropolis filter on the lifted path graph: Resampling and
laziness}
We now prepare the second step of our three-step procedure by lifting the
path graph. We recall that, for a lifting, the steady state $\pi_{\LIFTED{i
}{\pm}}$
of the lifted transition matrix $P$ must satisfy two conditions. First,
$\pi_{\LIFTED{i}{-}} + \pi_{\LIFTED{i}{+}} = \pi_i$.
Here, we choose
\begin{equation}
 \pi_{\LIFTED{i}{-}} =
 \pi_{\LIFTED{i}{+}} = \half \pi_i.
 \label{equ:LiftingOne}
\end{equation}
Second, the flows in the lifted Markov chain must reproduce those of the
collapsed Markov chain
\begin{equation}
\sum_{\sigma, \sigma' = \pm 1}
 \pi_{\LIFTED{i}{\sigma}} P_{\LIFTED{i}{\sigma},
 \LIFTED{j}{\sigma'}} =
 \pi_i P_{ij}
 \label{equ:LiftingTwo}
\end{equation}

\subsubsection{Non-reversibility parameter}
The following transition matrix is a lifting of
\eq{equ:TransitionMatrixGeneralCollapsed} with the restrictions of
\eqtwo{equ:LiftingOne}{equ:LiftingTwo}
\begin{equation}
 \begin{tikzcd}[column sep = 2.5cm, row sep = 0.5cm]
\circled{$\LIFTED{(i-1)}{+}$} \arrow{r}{(\half p + \delta)\minfunc{i}{i-1}}&
\circled{\;\;\;$\LIFTED{i}{+}$\;\;\;} \arrow[yshift=-1.0ex]{l}{(\half p -
\delta) \minfunc{i-1}{i}}
 \arrow[loop above]{}{\underbrace{\text{\small $1 - \Omega^+_i$}
} }
\arrow{r}{ (\half p + \delta) \minfunc{i+1}{i}}
\arrow[xshift=-1.0ex,swap]{d}{\Gamma_i^+ } &
\circled{$\LIFTED{(i+1)}{+}$} \arrow[yshift=-1.0ex]{l}{(\half p - \delta)
\minfunc{i}{i+1}}
\\
\circled{$\LIFTED{(i-1)}{-}$} \arrow[yshift=1.0ex]{r}{(\half p - \delta)
\minfunc{i}{i-1}}
&
\circled{\;\;\;$\LIFTED{i}{-}$\;\;\;} \arrow[yshift=1.0ex]{r}{(\half p - \delta)
\minfunc{i+1}{i}}
 \arrow[loop below]{}{\overbrace{\text{\small $1 -
\Omega^-_i$}}}
\arrow{l}{(\half p + \delta) \minfunc{i-1}{i}}
\arrow[xshift=1.0ex,swap]{u}{\Gamma_i^- }&
\circled{$\LIFTED{(i+1)}{-}$} \arrow{l}{(\half p  + \delta) \minfunc{i}{i+1}},
\end{tikzcd}
\label{equ:LiftedTransition}
\end{equation}
where we have introduced resampling rates $\Gamma_i^\pm$ and
laziness\footnote{A Markov chain that remains on
the same site is called \quot{lazy}~\cite{Levin2008}}
parameters $\Omega_i^{\pm}$.
This works for $\pi_{i^\pm} = \pi_i/2$ because in the combined flows from
$i^\pm $ to $(i+1)^\pm$, the terms in $\delta $ cancel exactly.
In order for \eq{equ:LiftedTransition} to describe a transition matrix, the
outgoing arrows from each lifted site must be
non-negative and sum up to one. This fixes the resampling
rates, from \eq{equ:LiftedTransition}, as
\begin{align}
 \Gamma_i^+ =  \Omega_i^+ -
 (\thalf p + \delta) \minfunc{i+1}{i}
- (\thalf p - \delta )\minfunc{i-1}{i} \label{equ:GammaiPlusFirst}, \\
 \Gamma_i^- =  \Omega_i^- - (\thalf p - \delta )
\minfunc{i+1}{i} - (\thalf p + \delta )\minfunc{i-1}{i}.
\label{equ:GammaiMinusFirst}
\end{align}
In addition, we impose the steady state $\pi_{i^\pm} = \pi_i/2$
by requiring the incoming flows to equal the
weights  $ \pi_i/2$:
\begin{align}
 \pi_i = (1 - \Omega_i^+) \pi_i + \Gamma_i^- \pi_i + (\thalf p -
\delta) \minb{\pi_i, \pi_{i+1}} + (\thalf p + \delta ) \minb{\pi_{i-1}, \pi_i},
\\
 \pi_i = (1 - \Omega_i^-) \pi_i + \Gamma_i^+ \pi_i + (\thalf p +
\delta) \minb{\pi_i, \pi_{i+1}} + (\thalf p - \delta ) \minb{\pi_{i-1}, \pi_i}.
\end{align}
Dividing by $\pi_i$ and rearranging gives
\begin{align}
\Gamma_i^+ =  \Omega_i^-   - (\thalf p +
\delta) \minfunc{i+1}{i} - (\thalf p - \delta ) \minfunc{i-1}{i},
\label{equ:Resampling1}
\\
 \Gamma_i^- =  \Omega_i^+   - (\thalf p -
\delta) \minfunc{i+1}{i} - (\thalf p + \delta ) \minfunc{i-1}{i}.
\label{equ:Resampling2}
\end{align}
In view of \eqtwo{equ:GammaiPlusFirst}{equ:GammaiMinusFirst},  the only
consistent choice
for the laziness parameters is $\Omega_i^\pm = \Omega_i$. We arrive at
\begin{align}
 \Gamma_i^+ &=  \Omega_i -
 (\thalf p + \delta) \minfunc{i+1}{i}
- (\thalf p - \delta )\minfunc{i-1}{i}, \label{equ:GammaiPlus} \\
 \Gamma_i^- &=  \Omega_i - (\thalf p - \delta )
\minfunc{i+1}{i} - (\thalf p + \delta )\minfunc{i-1}{i},
\label{equ:GammaiMinus}
\end{align}
which, if all probabilities are positive, produces a transition matrix
that satisfies global balance.
The \quot{skew} choice of $\Omega_i$ either cancels $\Gamma_i^+$
or $\Gamma_i^-$, leaving the other resampling probability positive
\begin{equation}
  \Omega_i^{\rm{skew}} =  \max\limits_{\sigma = \pm 1} \glc
              \glb \frac12 p + \delta \grb \minfunc{i+\sigma}{i} + \glb \frac12
p - \delta \grb
\minfunc{i-\sigma}{i} \grc.
\end{equation}
This choice provides additional simplifications beyond the choice $\delta=
p/2$ for the non-reversibility parameter, which leaves at most two outgoing
probability flows per lifted site and which will allow us to recursively compute
the \GM.

\subsection{Square-wave steady state}
\label{sec:SquareWave}

The square-wave steady state is defined by
\begin{equation}
\pi_i^\squarewave =
\begin{cases}
\frac{4}{3N} & \text{for $i \in \SET{2,4 \TO N}$ } \\
\frac{2}{3N}& \text{for $i \in  \SET{1,3\TO N-1}$}
\end{cases},
\label{equ:DHNSquareWave}
\end{equation}
where we suppose that $N$ is even: $N=2M$.
By an exact computation, we will show that the inverse gap, in other words the
relaxation time, scales as $\sim N^2$ on the path graph, and that on the lifted
path graph, we can reach a $\sim N$ relaxation time. The same results are
obtained for the \Kemeny, with generic scaling $\sim N^2$ on the path graph,
and
$\sim N$ on the lifted path graph. The speedup is thus as for the flat steady
state. However, this speed-up can likely not be reproduced within the framework
of \eq{equ:LiftedTransition} for more general steady states, for example a
$\pi$ with periodicity $3$ or larger.

\subsubsection{Square-wave steady state, path graph}
\label{sec:SquareWavePath}

The transition matrix on the path graph with the square-wave steady state is
given by
\begin{equation}
\begin{tikzcd}[column sep=1.2cm]
\circled{$2i-1$}
  \arrow[yshift=0.5ex]{r}{\half p}
&
\circled{\hphantom{0}$2i$\hphantom{0}}
  \arrow[loop above]{}
        {\underbrace{1-\half p}}
  \arrow[yshift=0.5ex]{r}{\frac14 p}
  \arrow[yshift=-0.5ex]{l}{\frac14 p}
&
\circled{$2i+1$}
  \arrow[loop above]{}
        {\underbrace{1- p}}
  \arrow[yshift=0.5ex]{r}{\frac12 p}
  \arrow[yshift=-0.5ex]{l}{\half p}
&
\circled{$2i+2$}
  \arrow[yshift=-0.5ex]{l}{\frac14 p}
\end{tikzcd} .
\label{equ:TransitionMatrixSquareCollapsed}
\end{equation}
The eigenvalue spectrum and the associated eigenvectors can be derived in closed
form using a modified Fourier basis, with the $k$th basis vector defined as
\begin{equation}
X_k(i)= \cosc{\frac{\pi k (i + \epsilon_k)}{N}}  \times
\begin{cases}
\alpha_k & \text{for $i \in \SET{2,4 \TO N}$} \\
\beta_k &  \text{for $i \in \SET{1,3\TO N-1}$} ,
\end{cases}
\end{equation}
with $k \in \SET{0 \TO M-1}$ and with  $\alpha_k$ and $\beta_k$
to be determined. The term $\epsilon_k$ is here
for
the boundary conditions (on sites $i=1$ and  $i=N$).
For $k \in \SET{1 \TO M-1}$,  $\epsilon_k$ depends on $k$, while for
$k=0$,
$\epsilon_k$ is singular. In the following, we compute eigenvalues
for the first case, while the $k=0$ term yields eigenvalues
$1$ and $1 - 3p/4$.

The one-step time evolution of the vector $X_k$, for $k \in
\SET{1 \TO M-1}$, yields
\begin{equation}
  Y_k(i) =
  \begin{cases}
     \frac{p}{2} X_k(i-1) + \frac{p}{2} X_k(i+1) + (1 - \frac{p}{2}) X_k(i) &
\text{for $i \in \SET{2, 4 \TO N}$} \\
    \frac{p}{4} X_k(i-1) + \frac{p}{4} X_k(i+1) + (1 - p) X_k(i) & \text{for $i \in
\SET{1, 3 \TO N-1}$},
  \end{cases}
\end{equation}
resulting in
\begin{equation}
  Y_k(i) = \cosc{ \frac{\pi k (i + \epsilon_k)}{N} }
  \begin{cases}
      \beta_k p \cosb{\frac{\pi k}{N}} + \alpha_k (1 - \frac{p}{2}) &
\text{for $i \in \SET{2, 4 \TO N}$} \\
      \alpha_k \frac{p}2 \cosb{\frac{\pi k}{N}} + \beta_k (1 - p) & \text{for
$i \in \SET{1, 3 \TO N-1}$}
  \end{cases} .
\end{equation}
We want $Y_k = \lambda X_k$, where $ \lambda$ are the eigenvalues of the
transition matrix. So, we consider the following $2\times 2$ matrix $A_k$:
\begin{equation}
A_k =
\begin{pmatrix}
1-\dfrac{p}{2} & p\cosa{\dfrac{\pi k}{N}}\\[1em]
\dfrac{p}{2}\cosa{\dfrac{\pi k}{N}} & 1-p
\end{pmatrix}\quad  \text{for $k \in \SET{1 \TO M-1}$},
\end{equation}
with eigenvalues
\begin{equation}
\lambda_\pm(k) =
  \frac14
  \glc  4 - 3p \pm p \sqrt{1 + 8 \cosb[2]{ {\frac{\pi k}{N}} }}
 \grc \quad
  \text{for $k \in \SET{1 \TO M-1}$} .
\label{equ:BulkEigenvaluesSquare}
\end{equation}
The complete spectrum is given by the $2(M-1)$ eigenvalues of
\eq{equ:BulkEigenvaluesSquare}, in addition to the isolated eigenvalues $1$ and
$1-3p/4$.
Checking the boundary conditions, we obtain
\begin{equation}
    \epsilon_k = \frac{1}{\theta_k}
    \arctan \glb \frac{\cos \theta_k - \dfrac{\alpha_k}{2 \beta_k}}{\sin
\theta_k} \grb \quad
    \text{with}\ \theta_k = \frac{\pi k}{N}.
\end{equation}
The second-largest eigenvalue is
\begin{equation}
\lambda_+(k=1) =
\glc 4-3p + p\sqrt{1 + 8 \cosb[2]{\pi/N}} \grc / 4
\sim 1 - p\frac{\pi^2}{3 N^2} + p\frac{\pi^4}{27 N^4} + ...
\end{equation}
so that the inverse gap is
\begin{equation}
\taurel^\squarewave = ( 1 - \lambda_2^\squarewave )^{-1} \sim \frac1p   \frac{3
N^2 }{\pi^2}.
\end{equation}
As for the flat steady state (see \eq{equ:GapUniformPath}),
the relaxation time scales as  $\sim N^2$. On a coarse-grained level, it
corresponds to diffusion on the path graph.

\subsubsection{Square-wave steady state, lifted path graph}

The square-wave steady state of \eq{equ:DHNSquareWave} satisfies
$\pi_{i-1}^\squarewave = \pi_{i+1}^\squarewave$ for sites $i$
in the interior of the path graph so that we have
$\minfunctext{i-1}{i} = \minfunctext{i+1}{i}$.
\footnote{From now on we write $\pi$
instead of $\pi^\squarewave$, for ease of notation.}
On the lifted path graph,
the resampling probabilities $\Gamma_i^+ = \Gamma_i^-$
of \eqtwo{equ:Resampling1}{equ:Resampling2} are thus the same, and we may
choose them arbitrarily small and, in particular, of order $ \epsilon \sim
1/N$. With a non-reversibility $0 \le \delta \le p/2$,
we have
an irreducible transition matrix
with a square-wave steady state that is aperiodic even for $\epsilon=0$:
\begin{equation}
 \begin{tikzcd}[column sep = 1.2cm, row sep = 0.8cm]
 \circled{$1+$} \arrow{r}{\half p + \delta}
 \arrow[loop above]{}{\underbrace{\text{\small $1 - p - \epsilon$}}}
\arrow[xshift=-0.5ex,swap]{d}{\half p + \epsilon - \delta}
 &
\circled{$2+$} \arrow{r}{\quarter p + \half \delta}
\arrow[yshift=-1.0ex]{l}{\quarter p - \half \delta}
 \arrow[loop above]{}{\underbrace{\text{\small $1 - \thalf p - \epsilon$}}}
\arrow[xshift=-0.5ex,swap]{d}{\epsilon }
&
\circled{$3+$}
\arrow[yshift=-1.0ex]{l}{\half p - \delta}
 \arrow[loop above]{}{\underbrace{\text{\small $1 - p - \epsilon$}}}
\arrow{r}{ \half p + \delta}
\arrow[xshift=-0.5ex,swap]{d}{\epsilon } &
\circled{$4+$} \arrow[yshift=-1.0ex]{l}{\quarter p - \half \delta}
 \arrow[loop above]{}{\underbrace{\text{\small $1 - \thalf p - \epsilon$}}}
\arrow[xshift=-0.5ex,swap]{d}{\quarter p + \epsilon + \half \delta}
\\
\circled{$1-$}\arrow[yshift=1.0ex]{r}{\half p - \delta}
 \arrow[loop below]{}{\overbrace{\text{\small $1 - p - \epsilon$}}}
\arrow[xshift=0.5ex,swap]{u}{\half p + \epsilon + \delta}
&
\circled{$2-$} \arrow[yshift=1.0ex]{r}{\quarter p - \half \delta}
\arrow{l}{\quarter p + \half \delta}
 \arrow[loop below]{}{\overbrace{\text{\small $1 - \thalf p - \epsilon$}}}
\arrow[xshift=0.5ex,swap]{u}{\epsilon}
&
\circled{$3-$} \arrow[yshift=1.0ex]{r}{\half p - \delta}
 \arrow[loop below]{}{\overbrace{\text{\small $1 - p - \epsilon$}}}
\arrow{l}{\half p + \delta}
\arrow[xshift=0.5ex,swap]{u}{\epsilon }
&
\circled{$4-$} \arrow{l}{\quarter p + \half \delta}
 \arrow[loop below]{}{\overbrace{\text{\small $1 - \thalf p - \epsilon$}}}
\arrow[xshift=0.5ex,swap]{u}{\quarter p + \epsilon - \half \delta}
\end{tikzcd} .
\label{equ:LiftedTransitionNonReversibleSquare}
\end{equation}
This solution with small  resampling rates in the interior of the lifted path
graph is special to the square-wave steady state. With a larger period,
say, with $\pi_i = \pi_{i\pm 3}$, arbitrarily small resampling rates
can then not be realized and liftings of the type we consider here then
produce no appreciable speedup.
The scenario of pairs of real eigenvalues merging and separating into the
complex plane is again realized (see \fig{fig:Lifted_Square}).

\subsubsection{\Kemeny, square-wave steady state, path graph}

For an even number of sites $N=2M$, we have the following quantities for the
square-wave steady state:
\begin{align}
  & \pi_i =
  \begin{cases}
    \frac{4}{3N} &  \text{for $i \in \SET{2, 4 \TO N}$} \\
    \frac{2}{3N} & \text{for $i \in \SET{1, 3 \TO N-1}$ } 
  \end{cases},  \\
  & P_{i,i+1} =
  \begin{cases}
    \frac{p}{4} &  \text{for $i \in \SET{2, 4 \TO N-2}$} \\
    \frac{p}{2} & \text{for $i \in \SET{1, 3 \TO N-1}$}  \\
    0 & \text{for $i=N$}
  \end{cases}, \\
  & \Pi_i = \sum_{k=1}^{i} \pi_k =
  \begin{cases}
    \frac{i}{N} &  \text{for $i \in \SET{2, 4 \TO N}$} \\
    \frac{2}{3N} + \frac{i-1}{N} & \text{for $i \in \SET{1,3 \TO N-1}$}
  \end{cases}.
\end{align}
We use \eq{equ:tausol} for the \MFPT:
\begin{equation}
 \tau_{1k} = \sum_{i=1}^{k-1} \frac{1}{\pi_i P_{i,i+1}} \Pi_i.
\end{equation}
With $ \taukemeny = \sum_{k=1}^{N} \tau_{1k}  \pi_k $ (see
\eq{equ:KemenySpectrumMFPT}), we arrive at
\begin{equation}
 \taukemeny = \sum_{i=1}^{N-1} \sum_{k=i+1}^{N} \frac{1}{\pi_i P_{i,i+1}} \Pi_i
\pi_k
 =\sum_{i=1}^{N-1} \frac{1}{\pi_i P_{i,i+1}} \Pi_i (1 - \Pi_i)
  = \sum_{k=1}^{N} A_k,
  \label{equ:SumAk}
\end{equation}
where $A_k$ is given by
\begin{equation}
  A_k = \begin{cases}
    \frac{3k(2M-k)}{2Mp} &  \text{for $k \in \SET{2, 4 \TO N}$} \\
    \frac{(6M-3k+1)(3k-1)}{6Mp} & \text{for $k \in \SET{1, 3 \TO N-1}$}
  \end{cases} .
\end{equation}
The sum in \eq{equ:SumAk} yields the \Kemeny,
\begin{equation}
  \taukemeny = \frac{2(3M^2-1)}{3p},
\end{equation}
which is confirmed by our \ExactDiag.

\subsubsection{\Kemeny, square-wave steady state, lifted path graph}

\begin{figure}[htb]
	\centering
\includegraphics[width=\columnwidth]{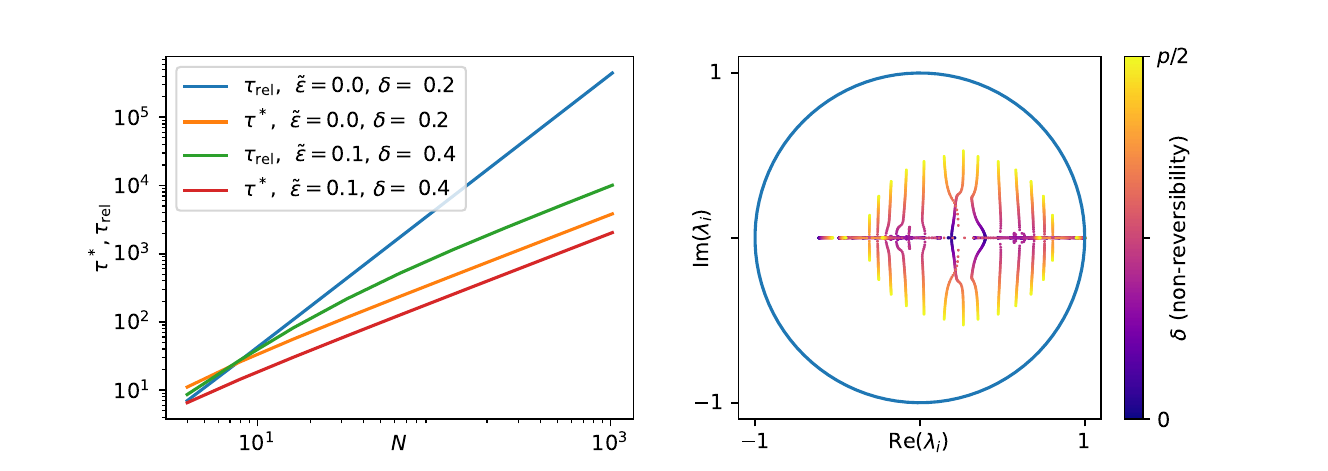}
  \caption{Lifted path graph with square-wave steady state
(from \ExactDiagShort).
  \subcap{a} Relaxation time $\taurel$ (inverse gap)
  and \Kemeny $\taukemeny$
as a function of $N$ for $p=0.8$ for different non-reversibilities
$\delta$ and resampling rates $\epsilon = \epsilontilde / N$.
The \Kemenys for $\delta = p/2$ agree with  \eq{equ:KemenyLiftedSquare}.
For $\epsilon=0$, we have $\taurel \sim N^2$, yet $\taukemeny \sim N$.
  \subcap{b} Spectrum of the transition matrix for $N=16$, $p=0.8$ as a
function of the non-reversibility $\delta$.
}
\label{fig:Lifted_Square}
\end{figure}

We consider the case $\delta = p/2$ of maximum non-reversibility,
\begin{equation}
 \begin{tikzcd}[column sep = 1.2cm, row sep = 0.8cm]
 \circled{$1+$} \arrow{r}{p }
 \arrow[loop above]{}{\underbrace{\text{\small $1 - p - \epsilon$}}}
\arrow[xshift=-0.5ex,swap]{d}{\epsilon}
 &
\circled{$2+$} \arrow{r}{\half p }
 \arrow[loop above]{}{\underbrace{\text{\small $1 - \thalf p - \epsilon$}}}
\arrow[xshift=-0.5ex,swap]{d}{\epsilon }
&
\circled{$3+$}
 \arrow[loop above]{}{\underbrace{\text{\small $1 - p - \epsilon$}}}
\arrow{r}{ p }
\arrow[xshift=-0.5ex,swap]{d}{\epsilon } &
\circled{$4+$}
 \arrow[loop above]{}{\underbrace{\text{\small $1 - \thalf p - \epsilon$}}}
\arrow[xshift=-0.5ex,swap]{d}{\half p + \epsilon }
\\
\circled{$1-$}
 \arrow[loop below]{}{\overbrace{\text{\small $1 - p - \epsilon$}}}
\arrow[xshift=0.5ex,swap]{u}{p + \epsilon}
&
\circled{$2-$}
\arrow{l}{\half p}
 \arrow[loop below]{}{\overbrace{\text{\small $1 - \thalf p - \epsilon$}}}
\arrow[xshift=0.5ex,swap]{u}{\epsilon}
&
\circled{$3-$}
 \arrow[loop below]{}{\overbrace{\text{\small $1 - p - \epsilon$}}}
\arrow{l}{p}
\arrow[xshift=0.5ex,swap]{u}{\epsilon }
&
\circled{$4-$} \arrow{l}{\half p}
 \arrow[loop below]{}{\overbrace{\text{\small $1 - \thalf p - \epsilon$}}}
\arrow[xshift=0.5ex,swap]{u}{\epsilon }
\end{tikzcd},
\label{equ:LiftedTransitionNonReversibleSquare2}
\end{equation}
with laziness parameters
\begin{equation}
 \Omega_i = \begin{cases}
    \frac{p}{2} + \epsilon &  \text{for  $i \in  \SET{2, 4, \TO N}$ } \\
    p + \epsilon & \text{for $i \in \SET{1,3 \TO N-1}$ }
  \end{cases} .
\end{equation}
To compute the \Kemeny on the lifted path graph for the
square-wave steady state, we use
\begin{align}
        \tau_{1k}^{++} & = \sum_{i=1}^{k-1} \frac1{\Omega_i}\glb 1 +
\frac{\Omega_i -2 P_{i,i+1}}{\pi_i P_{i,i+1}} \Pi_i \grb,  \\
     \sum_{k=1}^{N} \tau_{1k}^{++} \pi_k & = \sum_{i=1}^{N} \frac1{\Omega_i} \glb  1 +
\frac{\Omega_i -2 P_{i,i+1}}{\pi_i P_{i,i+1}} \Pi_i \grb(1 - \Pi_i) =
\sum_{k=1}^N B_k,
\label{equ:SumBk}
\end{align}
where $B_k$ is given by
\begin{equation}
  B_k = \begin{cases}
    \dfrac{(2M-k)(3 \epsilon k +p)}{Mp (2 \epsilon +p)} &
        \text{for $k \in \SET{2, 4 \TO N}$} \\
    \dfrac{(1-3k+6M)(p + \epsilon (3k-1))}{6Mp} &
    \text{for $k \in \SET{1, 3 \TO N-1}$}
  \end{cases} .
\end{equation}
The right-most sum  in \eq{equ:mftpsmaller} of \app{app:formula_kemeny_lifted}
yields
\begin{equation}
  \sum_{k=1}^{N} \tau_{1k}^{++} \pi_k  =
\frac{1}{6}\glc\frac{4(-1+3M^2)}{p}+\frac{-1+3M-6M^2}{\epsilon+p}-\frac{6(-1+M)
M}{2\epsilon+p} \grc.
\label{equ:lifted_squarewave_sum1}
\end{equation}
There is also a second term (for $k \neq 1$)
\begin{equation}
\tau_{1k}^{+-} = \tau_{1k}^{++} + \frac1{\Omega_k} \glc 1 +
\frac2{\pi_k}(1-\Pi_k) \grc =  \tau_{1k}^{++} + C_k.
\end{equation}
We use
\begin{equation}
  C_k \pi_k =
  \begin{cases}
    \dfrac{2(2-3k+6M)}{3Mp + 6 \epsilon M} &  \text{for $k \in \SET{2, 4 \TO
N}$} \\
    \dfrac{2-3k+6M}{3Mp + 3 \epsilon M} & \text{for $k \in \SET{3, 5 \TO N-1}$}
  \end{cases}
\end{equation}
to compute
\begin{equation}
   \sum_{k=2}^{N} C_k \pi_k = 
\frac{1}{3}\glc\frac{(-1+M)(-1+3M)}{M(\epsilon+p)}+\frac{-2+6M}{2\epsilon+p}
\grc.
  \label{equ:lifted_squarewave_sum2}
\end{equation}
There
remains one term, from \eq{equ:mftpcoincidingpm} (see
\app{app:formula_kemeny_lifted}),
\begin{align}
    & \tau_{11}^{+-} = \frac1{\Omega_1}\glc  1 + \frac2{\pi_1}(1- \pi_1)\grc,
\\
    & \tau_{11}^{+-} \pi_1 = \frac{6M-1}{3M(p+ \epsilon)}.
\label{equ:lifted_squarewave_sum3}
\end{align}
Because, by construction, $\pi_{i,\sigma} = \frac12 \pi_i$, the \Kemeny is
the sum of
\eq{equ:lifted_squarewave_sum1} and one half of
\eqtwo{equ:lifted_squarewave_sum2}{equ:lifted_squarewave_sum3}, so that we
obtain
\begin{equation}
   \taukemeny =
\frac{8\epsilon^2(-1+3M^2)+6\epsilon(-2+4M+3M^2)p+(-5+18M)p^2}{6p(\epsilon+p)
(2\epsilon+p)}.
\label{equ:KemenyLiftedSquare}
\end{equation}
This formula is confirmed by \ExactDiag (see \subfig{fig:Lifted_Square}{a}). For
$\epsilon = \epsilontilde / N$, we have
\begin{equation}
  \taukemeny \sim M\ \frac{3 (\epsilontilde + 2p)}{2 p^2},
\end{equation}
which continues to scale as $\taukemeny\sim N$ for $\epsilontilde =
0$ where the inverse gap scales as $\taurel \sim N^2$.

\subsection{\Kemenys and inverse gaps for \WEDGE and \V steady states}
\label{sec:WedgeVshape}

In this section, we analytically determine the \Kemeny for the \WEDGE steady
state on the path
graph and the lifted path graph and compare them to the relaxation times from
\ExactDiagShort. We also determine first-passage times $\tau_{1m}$
and $\tau_{1m}^{++}$ between the two edges of the path graph and the lifted path
graph. In addition, we follow the same program for the \V steady state on both
graphs. The model was previously studied in \REF{Hildebrand2004}.

In order to compare the two steady states, we use $N=2M$, and define the \WEDGE
only in the interval $\SET{1\TO M}$, where $\pi^\wedgeshape$ is twice as big
as $\pi^\Vshape$, while $\pi^\Vshape$ is defined for $\SET{1 \TO 2M = N}$:
\begin{align}
\pi_{i}^\wedgeshape & = \frac{8}{N^2} \glb  \frac{N+1}{2} -i \grb
=\frac{2}{M^2}\glb M + \frac12 -i \grb  \quad \text{for $i \in \SET{1 \TO
M}$},
\label{equ:PiWedgeDef}
\\
    \pi_i^\Vshape & = \frac4{N^2} \gle \frac{N+1}2 -i  \gre\quad \text{for $i
\in \SET{1 \TO 2M = N}$}.
\label{equ:PiVDef}
\end{align}
We again pad these distributions with phantom sites  $\pi_0^{\wedgeshape,
\Vshape} = 0$ as well as $\pi_{M+1}^\wedgeshape =  0$ and
$\pi_{2M+1}^\Vshape =  0$.
Inside the interval $i \in \SET{1, M-1}$ , the two transition matrices are
identical. On the lifted path graph, we use laziness parameters
\begin{align}
\Omega_i^\wedgeshape & = p + \epsilon \quad \text{for $i \in \SET{1 \TO M}$},
\label{equ:LazinessWedge}
\\
\Omega_i^\Vshape & = p +\epsilon \quad \text{for $i \in \SET{1 \TO 2 M}$}
\label{equ:LazinessV}
\end{align}
For our analytic calculations, we set $\epsilon=0$ unlike what we did for
the flat and square-wave steady states, as this parameter no longer serves to
break long cycles.
The following quantities are used:
\begin{equation}
P_{i,i+1}^\wedgeshape =
\begin{cases}
\dfrac{p}{2} \dfrac{M -i - \frac12}{M - i +
\frac12} & \text{for $i \in \SET{1, \dots, M-1}$}\\
0        &  \text{for $i=M$}
\end{cases},
\end{equation}
\begin{equation}
\Pi_i^\wedgeshape = \sum_{k \leq i} \pi_k^\wedgeshape = \frac{i(2M-i)}{M^2},
\end{equation}
\begin{equation}
P_{i,i+1}^\Vshape =
\begin{cases}
\frac{p}{2} \frac{M -i - \frac12}{M - i +
\frac12} & \text{for $i \in \SET{1, \dots, M-1}$}\\
\frac{p}2        &  \text{for $i \in \SET{M, \dots, N-1}$} \\
0 & \text{for $i =N$}
\end{cases},
\label{equ:P_Vshape}
\end{equation}
\begin{equation}
\pi_i^\Vshape P_{i,i+1}^\Vshape =
\begin{cases}
p \frac{N-2i-1}{N^2} & \text{for $i \in \SET{1, \dots, M-1}$}\\
\frac{p}{N^2}        &  \text{for $i =M$} \\
p \frac{2i-N-1}{N^2} & \text{for $i \in \SET{M+1, \dots, N-1}$}   \\
0 & \text{for $i =N$}
\end{cases},
\label{equ:pi_P_Vshape}
\end{equation}
\begin{equation}
\Pi_i^\Vshape  = \sum_{k \leq i} \pi_k^\Vshape =
\begin{cases}
\frac{2i(N-i)}{N^2} & \text{for $i \in \SET{1, \dots, M-1}$}\\
\frac12       &  \text{for $i =M$} \\
\frac12 + 2 \frac{(i - N/2)^2}{N^2} & \text{for $i \in \SET{M+1, \dots, N}$}  
\end{cases}.
\label{equ:PI_Vshape}
\end{equation}

We show that the \Kemeny for the \WEDGE scales as $\sim M^2$ for large
$M$ on the path graph, but as $\sim M$ for the the lifted path graph,
a behavior that we also find for the inverse gap. This
resembles the case of the flat steady state, but the case
$\epsilon=0$ is no longer exceptional. \footnote{For ease of notation,
we again write $\pi$ instead of $\pi^\wedgeshape$, \etcp.}

\subsubsection{\Kemeny, \WEDGE steady state, path graph}
\label{sec:computewedge}

With the \MFPT of  \eq{equ:tausol}, we have
\begin{equation}
 \tau_{1k} = \sum_{i=1}^{k-1} \frac{1}{\pi_i P_{i,i+1}} \Pi_i .
\end{equation}
Using $ \taukemeny = \sum_{k=1}^{M} \tau_{1k}  \pi_k $ (see
\eq{equ:KemenySpectrumMFPT}), we obtain
\begin{equation}
 \taukemeny = \sum_{i=1}^{M-1} \sum_{k=i+1}^{M} \frac{1}{\pi_i P_{i,i+1}} \Pi_i
\pi_k
 =\sum_{i=1}^{M-1} \frac{1}{\pi_i P_{i,i+1}} \Pi_i (1 - \Pi_i)
  = \sum_{k=1}^{M-1} \underbrace{\frac{2k(2M-k)(M-k)^2}{M^2p
(2M-2k-1)}}_{A_k}.
\end{equation}
Setting $X_k=M-k-\frac12$, we obtain
\begin{equation}
    A_k = \frac{-X_k^3}{p M^2} + \frac{-2X_k^2}{p M^2} + \frac{M^2 - \frac32}{p
M^2}X_k + \frac{M^2 -\frac12}{p M^2} + \frac{4 M^2-1}{16 p M^2}\frac1{X_k}.
\end{equation}
Isolating the $\frac1{X_k}$ term, replacing $X_k$ by $M-k-\frac12$, and
summing over $k$, we obtain
\begin{equation}
    \taukemeny = \frac{1}{p} \glb  \frac14 M^2 + \frac13 M - \frac18
-\frac{1}{12
M} \grb
+ \frac1p \glb \frac14 -  \frac{1}{16 M^2} \grb \sum_{k=1}^{M}
\frac{1}{M-k-\frac12}.
\end{equation}
This can be evaluated exactly. For large $M$, we have
\begin{align}
      \taukemeny &= \frac{1}{p} \glb \frac14 M^2 + \frac13 M - \frac18
-\frac{1}{12M} \grb \notag\\
&+ \frac1p \glb\frac14 -  \frac{1}{16 M^2}\grb \glc \log M + \gamma + 2 \log 2
-2 -
\frac1M +\bigOb{\frac1{M^2}} \grc \\
     &  = \frac{1}{p} \glc \frac14 M^2 + \frac13 M + \frac14 \log M +
\glb \frac{\log 2}{2} - \frac58 +\frac{\gamma}{4} \grb  \grc + \bigOb{
\frac1M }.
\end{align}

\subsubsection{\Kemeny, \WEDGE steady state, lifted path graph}

\begin{figure}[htb]
	\centering
\includegraphics[width=\columnwidth]{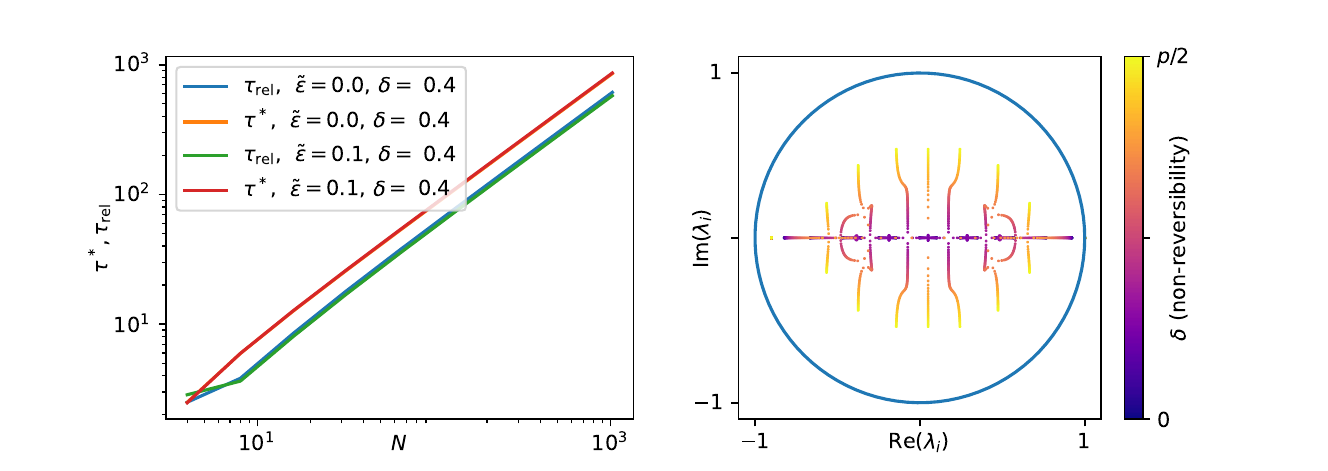}
  \caption{Lifted path graph with \WEDGE steady state
(from \ExactDiagShort, $p=0.8$).
  \subcap{a} Relaxation time $\taurel$ (inverse gap)
  and \Kemeny $\taukemeny$
as a function of $M$ for $p=0.8$ for different non-reversibilities
$\delta$ and resampling rates $\epsilon = \epsilontilde / M$.
The \Kemenys $\taukemeny \sim N$ for  $\delta = p/2$ and $\epsilon = 0 $ agree
with \eq{equ:Kemeny_Wedge_Lifted}.
\subcap{b} Spectrum of the transition matrix for $M=8$, $p=0.8$, as a
function of the non-reversibility $\delta$.
}
\label{fig:Lifted_Wedge}
\end{figure}

As, for simplicity, we set $\epsilon=0$ in \eq{equ:LazinessWedge}, one of  the
resampling rates $\Gamma_{\pm}$ is zero. In order to compute the \Kemeny, we
evaluate several terms in the sums of \app{app:formula_kemeny_lifted}. First, we
have
\begin{align}
      \tau_{1k}^{++} & = \sum_{i=1}^{k-1} \frac1{\Omega_i}\glb 1 +
\frac{\Omega_i -2 P_{i,i+1}}{\pi_i P_{i,i+1}} \Pi_i \grb \\
     \sum_{k=1}^{M} \tau_{1k}^{++} \pi_k & = \sum_{i=1}^{M} \frac1{\Omega_i} \glb  1 +
\frac{\Omega_i -2 P_{i,i+1}}{\pi_i P_{i,i+1}} \Pi_i \grb(1 - \Pi_i) \\
     & = \sum_{i=1}^{M} \frac{(i-M)^2 (4M^2 -1)}{(4(i-M)^2
-1) \cdot M^2 p} \label{equ:TelescopicSumWedge} \\
     & = \frac{1}{M^2 p} \sum_{i=1}^{M} \glc \glb M^2 -
\frac14 \grb + \glb \frac14 M^2 - \frac{1}{16} \grb \glb  \frac1{i-M-
\frac12} - \frac1{i-M + \frac12} \grb     \grc \\
     & = \frac{1}{M^2 p} \glc M \glb M^2 - \frac14 \grb
-\glb \frac14 M^2 - \frac{1}{16} \grb \glb 2 + \frac{1}{M- \frac12}
\grb  \grc  \\
    & = \frac{1}{p}\glb  M - \frac12 - \frac1{2M} \grb.
\label{equ:wedge_lifted_1}
\end{align}
In \eq{equ:TelescopicSumWedge}, the sum $\sum_{i=1}^{M} \glb
\frac1{i-M- \frac12} - \frac1{i-M + \frac12} \grb$ is telescopic.
There is also a second type of term (for $k \neq 1$), for which we compute the
sum weighted by $\pi_k$
\begin{equation}
\tau_{1k}^{+-} = \tau_{1k}^{++} + \frac1{\Omega_k} \glc 1 +
\frac2{\pi_k}(1-\Pi_k)\grc =  \tau_{1k}^{++} + C_k
\end{equation}
and, furthermore,
\begin{multline}
 \sum_{k=2}^{M} C_k \pi_k = \sum_{k=2}^{M} \frac1{\Omega_k} \glc 1 +
\frac2{\pi_k}(1-\Pi_k) \grc \pi_k \\
 = \sum_{k=2}^{M} \frac{1 -2k +2k^2 +2M -4kM +2M^2}{M^2 p }
  = \frac{2M^3 -6M^2 +7M-3}{3M^2p}.
\label{equ:wedge_lifted_2}
\end{multline}
\Eqq{equ:mftpcoincidingpm} (in \app{app:formula_kemeny_lifted})
contains one final term:
\begin{align}
    & \tau_{11}^{+-} = \frac1{\Omega_1}\glc  1 + \frac2{\pi_1}(1- \pi_1)\grc \\
    & \tau_{11}^{+-} \pi_1 = \frac1{p} \frac{2M^2 -2M +1}{M^2}.
    \label{equ:wedge_lifted_3}
\end{align}
Taking into account that the probability of each state $(k,\sigma)$ is
$\frac12\pi_k$, we compute the \Kemeny by adding \eq{equ:wedge_lifted_1} and
one-half of
\eqtwo{equ:wedge_lifted_2}{equ:wedge_lifted_3}:
\begin{equation}
    \taukemeny = \frac1p \glc  \frac43 M - \frac12 - \frac1{3M} \grc.
\label{equ:Kemeny_Wedge_Lifted}
\end{equation}
This result agrees with our \ExactDiag (see \fig{fig:Lifted_Wedge}).

`

\subsubsection{\Kemeny, \V steady state, path graph}

The \MFPT on the path graph is given by
\begin{equation}
    \tau_{1k} = \sum_{i=1}^{k-1} \frac{1}{\pi_i P_{i,i+1}} \Pi_i,
\end{equation}
(see \eq{equ:KemenyPath}).
\begin{equation}
   \taukemeny = \sum_{k=1}^{N} \tau_{1k}
   \pi_k = \sum_{k=1}^{N} \frac{1}{\pi_k P_{k,k+1}} \Pi_k (1 -
 \Pi_k).
\end{equation}
From \eqfromto{equ:P_Vshape}{equ:PI_Vshape}, we
obtain three contributions for the \Kemeny:
\begin{equation*}
    \taukemeny = \sum_{k=1}^{M-1} \underbrace{\frac{k(2M-k)(k^2 + 2M^2
-2kM)}{M^2
p(2M -2k -1)}}_{A_k} + \frac{M^2}{p} +\!\!\!\!\!
       \sum_{k=M+1}^{2M} \underbrace{\frac{k(2M-k)(k^2 + 2M^2 -2kM)}{M^2
p(2k -2M -1)}}_{B_k}.
\end{equation*}
Any term $A_k$ becomes, with $X_k = 2M-2k-1$,
\begin{equation}
    A_k = \frac{16 M^4 - (1+X_k)^4}{16 M^2 p X_k}.
    \label{equ:AkVshapeflat}
\end{equation}
We separate the $\frac1{X_k}$ contribution from the polynomial contribution in
$X_k$. Using $\sum_{k=1}^{M-1} \frac1{X_k} \sim \frac12( \log M + \gamma + 2
\log2 - \frac1{M} -\frac{11}{24M^2} )$ (with $\gamma$ the Euler--Mascheroni
constant), we obtain after summation
\begin{equation}
    \begin{aligned}
        \sum_{k=1}^{M-1} A_k = & \frac{16 M^4 - 1}{16 M^2 p} \frac12 \glc
\log M + \gamma + 2 \log 2 - \frac1{M} -\frac{11}{24M^2} + \bigOb{
\frac1{M^3}}  \grc\\
        &-\frac{(M-1)\left(3+M-2M^2+6M^3\right)}{48 M^2 p} .
    \end{aligned}
\end{equation}
For $B_k$, setting $Y_k= 2k-2M-1$, we obtain
\begin{equation}
    B_k = \frac{16 M^4 - (1+Y_k)^4}{16 M^2 p Y_k}.
\end{equation}
This is analogous to \eq{equ:AkVshapeflat}, with the replacement of
$X_k$ by
$Y_k$,
and with different bounds.
We obtain
\begin{multline}
     \sum_{k=M+1}^{2M} B_k =  \frac{16 M^4 - 1}{32 M^2 p}  \glc \log
M + \gamma + 2 \log 2 + \frac{1}{24M^2} + \bigOb{\frac1{M^4}} \grc \\
    -\frac{6 M^3 +16 M^2 +15M+8}{48 M p}.
\end{multline}
The complete formula is
\begin{multline}
     \taukemeny = \frac{36 M^4 - 8 M^3 -18M^2 -10M +3}{48 M^2 p} \\
     + \frac{16 M^4 -1}{16 M^2 p} \glc \log M + \gamma + 2 \log 2 -
\frac1{2M} -\frac{5}{24M^2} + \bigOb{\frac{1}{M^3}} \grc.
\end{multline}

Asymptotically, the behavior is
\begin{equation}
\taukemeny =
\frac{M^{2}}{p} \glb  \log M+\gamma+2\log 2 + \frac34 \grb
-
\frac{2M}{3p} - \frac{7}{12p} + \bigOb{\frac1M}.
\end{equation}

\subsubsection{\Kemeny, \V steady state, lifted path graph}

\begin{figure}[htb]
	\centering
\includegraphics[width=\columnwidth]{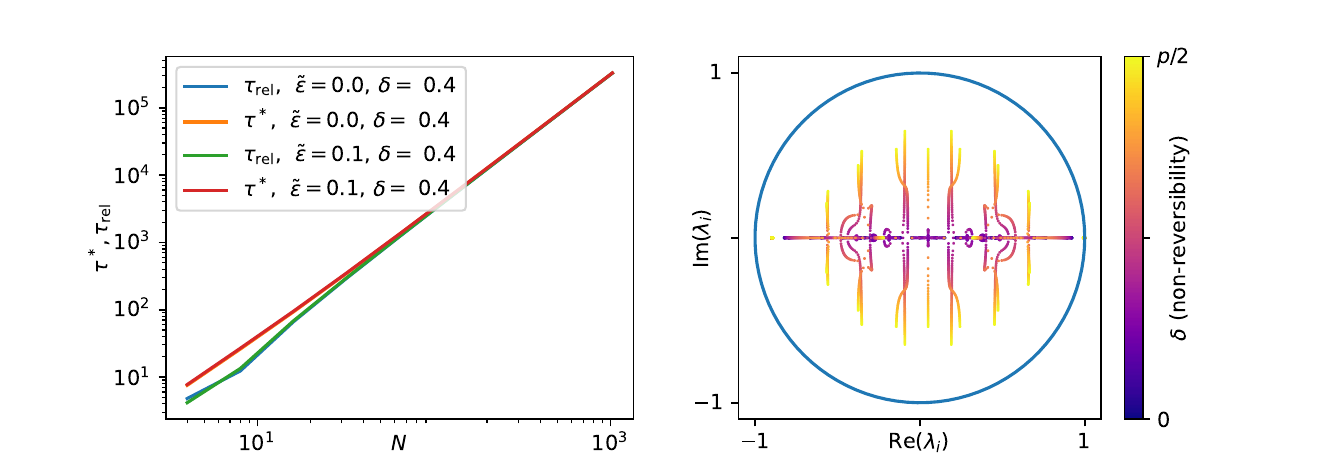}
  \caption{Lifted path graph with \V steady state
(from \ExactDiagShort, $p=0.8$).
  \subcap{a} Relaxation time $\taurel$ (inverse gap)
  and \Kemeny $\taukemeny$
as a function of $N$ for $p=0.8$ for different non-reversibilities
$\delta$ and resampling rates $\epsilon = \epsilontilde / N$.
The \Kemenys $\taukemeny \sim N^2$  for $\delta = p/2$ and $\epsilon = 0$ agree
with \eq{equ:Kemeny_V_Lifted}.
\subcap{b} Spectrum of the transition matrix for $N=16$, $\epsilon=0.15$ as a
function of the non-reversibility $\delta$.
}
\label{fig:Lifted_V}
\end{figure}

We evaluate expressions analogous to those of the \WEDGE steady state, with
appropriate values for $\pi_i$ and $P_{i,i+1}$.
To compute this \Kemeny, we start with
\begin{equation}
     \sum_{k=1}^{N} \tau_{1k}^{++} \pi_k =
     \sum_{k=1}^{N} \frac1{\Omega_k}\glb  1 +
\frac{\Omega_k -2 P_{k,k+1}}{\pi_k P_{k,k+1}} \Pi_k \grb (1 -\Pi_k),
\end{equation}
with \eqfromto{equ:P_Vshape}{equ:PI_Vshape}. The sum can be written as
\begin{equation*}
    \sum_{k=1}^{N} \tau_{1k}^{++} \pi_k = \sum_{k=1}^{M-1}
\underbrace{\frac{( k^2-2kM+2M^2)
(-1+4M^2)}{2(-1+4(k-M)^2) M ^ 2 p}}_{A_k}
 +
        \frac{1}{2p} + \sum_{k=M+1}^{N}\underbrace{ \frac{k(2M
-k)}{2M^2p}}_{B_k}.
\end{equation*}
To compute $A_k$, we note that
\begin{equation}
    \frac1{(2M-2k-1)(2M-2k+1)} = \frac12 \glb  \frac1{2M-2k-1} -
\frac1{2M-2k+1} \grb.
\end{equation}
We thus write $A_k = A_k^1 + A_k^2$, with
\begin{align}
     & A_k^1 =
\frac{(-1+4M^2)(2k^2-4kM+4M^2)}{8M^2(-1-2k+2M)
p }, \\
     & A_k^2 =
\frac{(1-4M^2)(2k^2-4kM+4M^2)}{8M^2(1-2k+2M)p}.
\end{align}
Setting $X_k=2M-2k-1$, we decompose $A_k^1$ into
polynomials in $X_k$, separate the $\frac1{X_k}$ term from the polynomial
contribution, then substitute $X_k$ by its expression in terms of $k$. We
obtain
\begin{align}
     A_k^1 &= \frac{(4M^2-1)(4M^2 + (1+X_k)^2)}{16M^2 p X_k}\\
     &= \frac{(4M^2 -1)(4M^2 +1)}{16 M^2 p} \frac{1}{2M-2k-1} +
           \frac{4M^2-1}{16 M^2 p}(2M-2k+1),
\end{align}
which gives for the sum
\begin{equation}
        \sum_{k=1}^{M-1} A_k^1 =  \frac{(4M^2 -1)(4M^2 +1)}{16 M^2 p}
           \sum_{k=1}^{M-1} \frac{1}{2M - 2k - 1}
            +  \frac{4M^2-1}{16 M^2 p} (M^2 -1).
\end{equation}
To compute $A_k^2$, setting $Y_k = 2M-2k+1$, we decompose $A_k^2$
into polynomials in $Y_k$, separate the $\frac1{Y_k}$ term from the polynomial
contribution, then substitute $Y_k$ by its expression in terms of $k$. We
obtain
\begin{align}
     A_k^2 &= \frac{(4M^2-1)(4M^2 + (-1+Y_k)^2)}{16M^2 p Y_k}\\
      &= -\frac{(4M^2 -1)(4M^2 +1)}{16 M^2 p} \frac{1}{2M-2k+1} -
                   \frac{4M^2-1}{16 M^2 p}(2M-2k-1),
\end{align}
leading to
\begin{multline}
      \sum_{k=1}^{M-1} A_k^2 =  -\frac{(4M^2 -1)(4M^2 +1)}{16 M^2 p}
              \sum_{k=1}^{M-1} \frac{1}{2M - 2k + 1} \\
  -  \frac{4M^2-1}{16 M^2 p} (M^2 -2M+1).
\end{multline}
Summing the terms $A_k^1$ and $A_k^2$  simplifies
nearly all terms of the form $\frac1{2M-2k \pm 1}$. Thus
\begin{equation}
    \sum_{k=1}^{M-1} \left( A_k^1 + A_k^2 \right ) = \frac{(M-1)
       (2M+1)^2}{4Mp}.
    \label{equ:A_k}
\end{equation}
The sum of the $B_k$ terms is, likewise,
\begin{equation}
    \sum_{k=M+1}^{2M} B_k = \frac{4M^2 -3M -1}{12 M p} .
    \label{equ:B_k}
\end{equation}
We also have the term
\begin{align}
    & \sum_{k=2}^{2M} \frac1{\Omega_k} \glc 1 + \frac2{\pi_k}(1-\Pi_k) \grc \pi_k  \\
    & = \sum_{k=2}^{M-1} \frac{4M^2 +M(2-4k) + 1-2k+2k^2 }{2 M^2p} + \frac{2
         + \frac{1}{M^2}}{2p} \\
    & \quad + \sum_{k=M+1}^{2M}\frac{ 2k(2M+1-k) -2M
-1}{2M^2 p} \\
    & = \frac{4M^3 -4M^2 +2M -1}{2 M^2 p }.
    \label{equ:C_k}
\end{align}
Finally, from \eq{equ:mftpsmaller}, the last term to compute the \Kemeny is
\begin{align}
    & \tau_{11}^{+-} = \frac1{\Omega_1}\glc  1 + \frac2{\pi_1}(1- \pi_1) \grc,
\\
    & \tau_{11}^{+-} \pi_1 = \frac1{p} \frac{4M^2 -2M+1}{2 M^2 }.
    \label{equ:D_k}
\end{align}
Remembering that $\pi_{k,\sigma} = \frac12 \pi_k$,
we can compute the \Kemeny by adding $\frac{1}{2p}$ to  \eqtwo{equ:A_k}{equ:B_k}
and to one half of \eqtwo{equ:C_k}{equ:D_k}. We then obtain
\begin{equation}
    \taukemeny = \frac1p \glb M^2 + \frac43 M - \frac12  - \frac1{3M} \grb.
\label{equ:Kemeny_V_Lifted}
\end{equation}
This result (which is valid for $\delta = p/2$ and $\epsilon=0$)
agrees with our \ExactDiag (see \fig{fig:Lifted_V}).

\subsection{\MFPTCAP for \WEDGE and \V steady states}

\label{sec:computemfpt}

The \MFPT $\tau_{1M}$ for the \WEDGE and the \V steady states on the path graph
are the same, as the transition matrices on the intermediate sites agree.
For the same reason, $\tau_{1M}^{++}$ also agrees for the two cases.

\subsubsection{\MFPTCAP, \WEDGE steady state, path graph}

We have, from \eq{equ:tausolAlternative}
\begin{equation}
     \tau_{1M} = \sum_{k=1}^{M-1} \frac{1}{\pi_k P_{k,k+1}} \Pi_k
      = \sum_{k=1}^{M-1} \underbrace{\frac{2k(2M-k)}{p(2M-2k-1)}}_{A_k}.
\end{equation}
Setting $X_k=2M-2k-1$, we write
\begin{equation}
    A_k = - \frac1p + \frac{4M^2-1}{2pX_k} - \frac{X_k}{2p}.
\end{equation}
All sums of the terms involving the $X_k$ can be evaluated, and  we obtain
\begin{equation}
     \tau_{1M} = -\frac{M^2 - 1}{2p} +\frac{4M^2-1}{4p} \glb \log M + 2\log 2
+ \gamma - \frac1M -\frac{11}{24M^2} \grb.
\end{equation}
Asymptotically, we have
\begin{multline}
        \tau_{1M} =  \frac1p M^2 \log M + \frac1p M^2 \glb  \gamma + 2\log 2 -
\frac12 \grb \\
         - \frac{1}{p}M - \frac1{4p} \log M  - \frac1{p}
         \glb \frac14 \gamma + \frac12 \log 2 - \frac1{24} \grb
+\bigOb{\frac{1}{M}}.
\end{multline}

\subsubsection{\MFPTCAP, \WEDGE steady state, lifted path graph}
\label{sec_MFPTWedgeLifted}
Let us compute $\tau_{1M}^{++}$. We have from \eq{equ:mftpsmaller}
\begin{equation}
     \tau_{1M}^{++} = \sum_{k=1}^{M-1} \underbrace{ \frac1{\Omega_k} \glb 1 +
\frac{\Omega_k -2 P_{k,k+1}}{\pi_k P_{k,k+1}} \Pi_k \grb }_{B_k}
    \label{equ:SumBktwo}
\end{equation}
with
\begin{equation}
    B_k = \frac{4M^2 -1}{p \glc 4 (M-k)^2 -1\grc} =
     \frac{4M^2 -1}{2p}  \glb \frac{1}{2M-2k-1} - \frac1{2M-2k+1} \grb.
\end{equation}
The sum in \eq{equ:SumBktwo} is thus telescopic, and evaluates to
\begin{equation}
    \tau_{1M}^{++} = \frac{2M^2 -M -1}{p}.
\end{equation}
Remarkably, the $\sim M^2 \log M $ contribution to the \MFPT on the path
graph has disappeared on the lifted path graph.

\section{Conclusions, outlook}
\label{sec:Conclusions}

In the present paper, we have studied Markov chains on a one-dimensional
lattice (the path graph) and their liftings, which effectively live on
one-dimensional ladders (the lifted path graph). On the path graph,
Markov chains are always reversible, and we can duplicate these
Markov chains onto the lifted path graph without many changes. In addition,
as we discussed, we can introduce a parameter that drives the
Markov chain gradually into non-reversibility. In the chosen example, we could
show how the qualitative differences between reversible and non-reversible
Markov chains build up gradually. We could also witness how the gap widens as a
result of the hybridization of different branches of the spectrum.
We complemented our exact calculations of the transition-matrix spectrum with a
Green's matrix approach, which gives access to the \MFPT and the \Kemeny. For
several choices of steady states, this allowed us to perform a detailed
analysis of
speedups through lifting, that was compared with  the results of
numerical
solutions. We expect the hybridization mechanism exposed in our
study of exactly solvable models to carry over to much more complicated
systems.

\section*{Conflict of interest}
The authors declare that they have no conflict of interest.

\section*{Acknowledgement}
W.K. acknowledges generous support by the Leverhulme Trust.

\appendix

\section{Examples of transition matrices}
\label{app:Examples}

In this appendix, we write out the schematic transition matrices of
\sect{sec:Introduction} and compute their spectrum.

\subsection{Eigenvalues and eigenvectors of transition matrix $P$, flat steady
state, path graph}
\label{app:ExampleTransitionPathFlat}

For the flat steady state on the path graph,
with the numbering scheme $1 = \circled{1}$, $2 = \circled{2}$,
\etc,  we have the transition matrix:
\begin{equation}
P =
\begin{bmatrix}
   1 - \half p &  \half p   & \oo & \oo \\
  \half p  &  1-p & \half p  & \oo  \\
 \oo & \half p  & 1 - p & \half p \\
 \oo & \oo & \half p  & 1- \half p   \\
\end{bmatrix} \ .
\label{equ:ExampleTransitionPathFlat}
\end{equation}
Here and in the following, the \quot{$\oo$}s stand for zeros. The transition
matrix $P$ of \eq{equ:ExampleTransitionPathFlat} is irreducible and it is
aperiodic by virtue of the presence of diagonal terms. It is doubly stochastic
(all rows and all columns sum to one), so that the steady state is
constant.
In addition to being doubly stochastic, it is symmetric, so it is reversible.

The (left) eigenvectors of \eq{equ:ExampleTransitionPathFlat} are
\begin{equation}
 e_h(i) = \cosc{\pi \glb i - \half \grb  \frac{h}{N} };\quad h=0 \TO
N-1;\quad
 i=1 \TO N.
\label{equ:EigenvectorFlatWalk}
\end{equation}
(the matrix is symmetric, and left and right eigenvectors agree, but this is
not the case for other transition matrices in the present paper).
We verify explicitly that, indeed, $e_h P = \lambda(h) e_h$.
For $i= 2 \TO N-1$, we set $\hat{e}_h = e_h P$ and obtain
\begin{align*}
  \hat{e}_h(i) &= \half p e_h(i+1) + \half p e_h(i-1) + (1-p) e_h(i) \\
               &= \half p \gld \cosc{\pi \glb i-\frac12 \grb \frac{h}{N}}
\cosb{\pi \frac{h}{N}} - \sinc{\pi \glb i-\half \grb \frac{h}{N}}
\sinb{\pi \frac{h}{N}} \grd \\
      &\quad + \half p \gld \cosc{\pi\glb i - \half \grb \frac{h}{N}}
    \cosb{\pi \frac{h}{N}} + \sinc{\pi \glb i - \half \grb  \frac{h }{N}}
\sinb{\pi \frac{h}{N}} \grd \\
      &\quad + (1-p) \cosc{ \pi \glb i - \half  \grb  \frac{h}{N}}
\\
   &= p \cosc{\pi \glb  i - \half \grb  \frac{h}{N}} \cosb{ \pi
        \frac{h}{N}} + (1-p) \cosc{ \pi \glb i - \half \grb  \frac{h}{N}} \\
  &= \underbrace{ \glc (1-p) + p \cosb{ \pi \frac{h}{N} }
        \grc}_{\lambda(h)\quad \text{(see \eq{equ:SpectrumCollapsed})}} e_h(i) .
\label{equ:EigenvectorFlatWalkComputation}
\end{align*}
Analogous  computations apply to the eigenvectors and eigenvalues for $i=1$ and
$i=N$, so that \eq{equ:SpectrumCollapsed} is established.

\subsection{Eigenvalues and eigenvectors of $P$, flat steady
state, lifted path graph}
\label{app:ExampleTransitionPathLiftedFlat}

To compute the transition-matrix spectrum for the flat steady state
on the lifted path graph, we use the numbering scheme
\begin{align*}
1  &=  \circled{1+} \quad
2  =  \circled{2+} \quad
3  =  \circled{3+} \quad
4  =  \circled{4+} \\
8  &=  \circled{1-} \quad
7  =  \circled{2-} \quad
6  =  \circled{3-} \quad
5  =  \circled{4-},
\end{align*}
here shown for $N=4$, where $(x,\pm ) \rightarrow (x,\mp)$ correspond
to $X \rightarrow 1-X$ when $x$ and  $x \pm 2 N$ are identified.
The transition matrix of \eq{equ:LiftedTransitionNonReversible} is then given by
\begin{equation}
P_{X,X'} = \glb \frac{p}{2} + \delta \grb \delta_{ X',X+1}  +
  \glb \frac{p}{2} - \delta \grb \delta_{X' ,X-1}
    + (1-p-\epsilon) \delta_{X',X} +  \epsilon \delta_{X',1-X}.
\end{equation}
We can explicitly write $P$, again for $N=4$, as
\begin{equation}
  P =
    \begin{bmatrix}
        1-p-\epsilon & \half p + \delta & \cdot & \cdot & \cdot & \cdot &
\cdot & \half p + \epsilon - \delta \\
        \half p - \delta & 1-p-\epsilon & \half p + \delta & \cdot & \cdot &
\cdot & \epsilon & \cdot \\
        \cdot & \half p - \delta & 1-p-\epsilon & \half p + \delta & \cdot &
\epsilon & \cdot & \cdot \\
        \cdot & \cdot & \half p - \delta & 1-p-\epsilon & \half p + \epsilon
+ \delta & \cdot & \cdot & \cdot \\
        \cdot & \cdot & \cdot & \half p +\epsilon - \delta & 1-p-\epsilon &
\half p + \delta & \cdot & \cdot \\ 
        \cdot & \cdot & \epsilon & \cdot & \half p - \delta & 1-p-\epsilon &
\half p + \delta & \cdot \\
        \cdot & \epsilon & \cdot & \cdot & \cdot & \half p - \delta &
1-p-\epsilon & \half p + \delta \\
        \half p + \epsilon + \delta & \cdot & \cdot & \cdot & \cdot & \cdot &
\half p - \delta & 1-p-\epsilon
    \end{bmatrix} .
\label{equ:TransitionMatrixFlatLiftedExplicit}
\end{equation}
This transition matrix is again doubly stochastic, so that the steady state is
indeed flat (and, in particular, independent of all parameters). For $\delta=0$,
it is symmetric, thus reversible, while it is non-reversible for $0<
|\delta|
< p/2$, with $p+\epsilon<1$, as in
\eq{equ:LiftedTransitionNonReversible}. In order to compute the eigenvalues, we
define
\begin{equation}
{\mathring \pi}_h(t)  =  \sum_{X=-N+1}^N \expb{i \theta  X}
\pi_X(t)\quad \text{with $ \theta = \pi h /N$}.
\label{equ:FourierBasisAppendix}
\end{equation}
The time-evolution equation for ${\mathring \pi}_h(t)$ is
\begin{equation}
    {\mathring \pi}_h(t+1) = \glc \glb \half p + \delta\grb
\expb{i\theta}
        + \glb\half p - \delta\grb \expb{-i\theta} +
(1-p-\epsilon)
        \grc {\mathring \pi}_h(t) + \epsilon \expb{i\theta} {\mathring
\pi}_{-h}(t).
\end{equation}
Together with an equivalent expression for ${\mathring \pi}_{-h}(t+1)$, this
yields two coupled linear equations represented by the matrix
\begin{equation*}
    A_h =
    \begin{bmatrix}
    \glb \half p + \delta \grb \expb{i\theta} + \glb \half p -
\delta \grb
        \expb{-i\theta} + 1 - p - \epsilon\hspace{-1cm} &
\epsilon \expb{i\theta} \\
    \epsilon \expb{-i\theta} & \hspace{-1cm} \glb \half p + \delta
\grb \expb{-i\theta} + \glb \half p -
        \delta \grb \expb{i\theta} + 1 - p - \epsilon
    \end{bmatrix} .
\end{equation*}
The eigenvalues of this complex-valued $2\times 2$ matrix are
\begin{equation}
    \lambda_{\pm}(\theta) = (1 - p - \epsilon) +
    p \cosa{\theta} \pm \sqrt{ \epsilon^2 - 4 \delta^2 \sina[2]{\theta}},
\end{equation}
which leads to the spectrum of \eqfromto{equ:Value0}{equ:LiftedValue_h} (see
\app{app:ComputerPrograms} for a Mathematica notebook).

\section{Transition matrix and \GM: Basic properties, Dirac notation}
\label{app:Dirac}
In this appendix, we review the representation of diagonalizable matrices, in
the context of the \GM (\sect{app:DiagonalizableMatrix}), and then connect the
\GM to the \MFPT (\sect{app:GM_MFPT}), deriving an expression that is used in
\sect{sec:GM_Kemeny}.

\subsection{Representation of diagonalizable matrices, \GM}
\label{app:DiagonalizableMatrix}

In this appendix, we illustrate general properties of diagonalizable
matrices in connection with the Dirac notation (see \app{app:ComputerPrograms}
for a Mathematica notebook and a Python program illustrating material in this
appendix). We also point out a notational subtlety that appears for a
complex-valued spectrum.

A diagonalizable transition matrix $P$, as any real-valued diagonalizable
$N\times N$  matrix, can be written as
\begin{align}
P &= R \Lambda L, \label{equ:RightLeft}\\
\intertext{where $\Lambda = \text{diag}(\lambda_1 \TO \lambda_N)$ is a diagonal
matrix. The transition matrix $P$ is real, but $R$, $\Lambda$, and $L$
are all possibly complex-valued. We have}
\ONE &= R L
\label{equ:RightLeftONE},
\end{align}
meaning that $L$ is the inverse of $R$. Associativity implies $\ONE = L R$ from
\eq{equ:RightLeftONE}, and it thus follows that
\begin{align}
LP &= \Lambda L, \label{equ:LeftEigenvectors}\\
PR &= R \Lambda.
\label{equ:RightEigenvectors}
\end{align}
On the right-hand side of \eq{equ:LeftEigenvectors},  row $m$ of $L$ is
multiplied with $\lambda_m$, and so is column $m$ of $R$ in
\eq{equ:RightEigenvectors}. Left eigenvectors $\SET{x_1 \TO x_N} $ with
eigenvalue $\lambda$ of the matrix $P$ satisfy $\lambda x^*_k =  \sum_l
x^*_l P_{lk}$, from which it follows that the rows of the  matrix $L$ in
\eq{equ:LeftEigenvectors} contain
the complex conjugates of the eigenvectors $L$ with their associated
eigenvalues.
Also, from \eq{equ:RightEigenvectors} it follows that the columns of $R$ contain
the $N$ right
eigenvectors of $P$.

We can write the matrices in \eq{equ:RightLeft} as $R = R_1 \PLUSPLUS R_N$
and $L = L_1 \PLUSPLUS L_N$, where $R_m$ is the matrix reduced to the  $m$th
column vector of $R$ (and is otherwise zero).
Likewise, $L_m$ is the matrix
reduced to the $m$th
row vector of $L$ (and is otherwise zero). It
follows that  $P R_m = \lambda_m R_m$, so that $R_m$ is
the matrix containing the $m$th right eigenvector on its $m$th column. It
also
follows that
$L_m P = \lambda_m L_m$, so that  (in view of the definition of
\eq{equ:DefEigenVectorLeft}) $L_m$ is the matrix containing the complex
conjugate of the $m$th left eigenvector on its $m$th row.
Therefore,
\begin{align}
P = \sum_m \lambda_m \KET{\R{m}} \BRA{\L{m}} \DiracCoord
P &= \glb R_1 \PLUSPLUS R_4 \grb \Lambda
\glb L_1 \PLUSPLUS L_4 \grb = \sum_m \lambda_m R_m L_m,
\label{equ:PDiracCoord}
\\
\ONE  =  \sum_m \KET{\R{m}} \BRA{\L{m}}
\DiracCoord
\ONE &= \glb R_1 \PLUSPLUS R_4 \grb
\glb L_1 \PLUSPLUS L_4 \grb = \sum_m R_m L_m,
\label{equ:OneDiracCoord}
\end{align}
as $R_m L_n  = \ZERO $, for $m \ne n$ ($R_m$ is non-zero only on
the $m$th column, and $L_n$ only on the $n$th row).
In \eq{equ:OneDiracCoord}, the normalization of the
eigenvectors is not fixed ($R_m$ could be replaced by $\alpha_m R_m$, if $L_m$
is replaced by $L_m / \alpha_m$).

Dirac notation, on the left-hand side of
\eqtwo{equ:PDiracCoord}{equ:OneDiracCoord}, consists in writing $R_m L_m$  as
the outer product
with the column vector $\KET{\R{m}} = \SET{\R{m}_1 \TO \R{m}_N}^T$ and the row
vector $\BRA{\L{m}} = \SET{\L{m}_1 \TO \L{m}_N}$.
\footnote{In this appendix we write $\L{m}_i$ for the complex conjugate of the
component $i$ of the $m$th eigenvector, in order to simplify the notation.
As discussed before, $L_m$ is the matrix containing the complex conjugate of
the $m$th left eigenvector.}
With
(for $N=4$)
\begin{equation}
\KET{1} \BRA{\pi}  \DiracCoord
 \begin{pmatrix}
 1 & \, . \, & \, . \, & \, . \, \\
 1 & . &.  & . \\
 1 & . &.  & . \\
 1 & . &.  & .
 \end{pmatrix}
 \begin{pmatrix}
 \pi_1 & \pi_2 & \pi_3 & \pi_N \\
 . & . &.  & . \\
 . & . &.  & . \\
 . & . &.  & .
 \end{pmatrix}
=
 \begin{pmatrix}
 \pi_1 & \pi_2 & \pi_3 & \pi_N \\
 \pi_1 & \pi_2 & \pi_3 & \pi_N \\
 \pi_1 & \pi_2 & \pi_3 & \pi_N \\
 \pi_1 & \pi_2 & \pi_3 & \pi_N
 \end{pmatrix},
 \label{equ:DiracOnePi}
\end{equation}
(where \quot{$.$} again indicate zeros) or, in other words, the matrix, with
elements
$\BRAN{x} \KET{1} \BRAN{\pi} \KET{y}$. Other example:
\begin{equation}
\KET{\R{2}} \BRA{\L{2}} \DiracCoord
 \begin{pmatrix}
\,.\, &  \R{2}_1 & \, . \, & \, . \, \\
 . & \R{2}_2 &.  & . \\
 . & \R{2}_3 &.  & . \\
 . & \R{2}_4 &.  & .
 \end{pmatrix}
 \begin{pmatrix}
 . & . &.  & . \\
 \L{2}_1 & \L{2}_2 & \L{2}_3  & \L{2}_4 \\
 . & . &.  & . \\
 . & . &.  & .
 \end{pmatrix}
=
 \begin{pmatrix}
 \R{2}_1 \L{2}_1 & \R{2}_1 \L{2}_2 & \R{2}_1 \L{2}_3  & \R{2}_1 \L{2}_4 \\
 \R{2}_2 \L{2}_1 & \R{2}_2 \L{2}_2 & \R{2}_2 \L{2}_3  & \R{2}_2 \L{2}_4 \\
 \R{2}_3 \L{2}_1 & \R{2}_3 \L{2}_2 & \R{2}_3 \L{2}_3  & \R{2}_3 \L{2}_4 \\
 \R{2}_4 \L{2}_1 & \R{2}_4 \L{2}_2 & \R{2}_4 \L{2}_3  & \R{2}_4 \L{2}_4
 \end{pmatrix},
 \label{equ:ExampleOuterProduct}
\end{equation}
Evidently, the traces of $\KET{1} \BRA{\pi}$, and $\KET{\R{2}} \BRA{\L{2}}$,
and more generally, of $\RMLM$ are all equal to one. As evident in
\eqtwo{equ:DiracOnePi}{equ:ExampleOuterProduct},
$\KET{\R{m}} \BRA{\L{m}}$ are rank-two tensors, that is, matrices.
In contrast, the inner product $L_m R_m$, written as  $\BRAN{\L{m}}
\KET{\R{m}}$, for example, for $m=2$,
\begin{equation}
\BRAN{\L{2}} \KET{\R{2}} \DiracCoord
L_2 R_2 =
 \begin{pmatrix}
 . & . &.  & . \\
 \L{2}_1 & \L{2}_2 & \L{2}_3  & \L{2}_4 \\
 . & . &.  & . \\
 . & . &.  & .
 \end{pmatrix}
 \begin{pmatrix}
\,.\, &  \R{2}_1 & \, . \, & \, . \, \\
 . & \R{2}_2 &.  & . \\
 . & \R{2}_3 &.  & . \\
 . & \R{2}_4 &.  & .
 \end{pmatrix}
=
 \begin{pmatrix}
.& .& .& . \\
.& 1& .& . \\
.& .& .& . \\
.& .& .& .
 \end{pmatrix},
\end{equation}
corresponds to a scalar.

From the spectral
representation of \eq{equ:PDiracCoord}, with \eq{equ:DiracOnePi}, we see that
\begin{equation}
 P^t
 = \KET{1} \BRA{\pi} + \sum_{m\ne 1} \lambda_m^t \RMLM,
\label{equ:chainspectral2}
\end{equation}
so that the \GM is
\begin{equation}
G = \sum_{t=0}^\infty \Delta_t = \sum_{m \neq 1} \frac{1}{1 - \lambda_m}
\KET{\R{m}} \BRA{\L{m}}.
\end{equation}
The two orthogonality conditions of \eqtwo{equ:LeftGorthog}{equ:Gorthog} follow
from this equation, as the sum on its right-hand side involves neither
$\KET{\R{1}}$ nor $\BRA{\L{1}}$.
As discussed, the trace of $\RMLM$ is one, so that we obtain
\begin{equation}
\Tr\ G = \sum_{t=0}^\infty \Tr\ \Delta_t = \sum_{m \neq 1} \frac{1}{1 -
\lambda_m},
\end{equation}
in other words a direct connection of the \GM (on the left) with the spectrum.
The expression, strictly speaking, is valid for a diagonalizable irreducible
transition matrix that need not be aperiodic, but by continuity it carries over
to the isolated points in parameter space where it is non-diagonalizable.
We will not discuss the non-diagonalizable case in the present paper, as it
adds no new features.

\subsection{Relation between the \GM and the \MFPT}
\label{app:GM_MFPT}

In this appendix, we derive \eq{equ:GFirstPassage}, that is, the representation
of \GUM elements in terms of the \MFPT. For a reminder, we
have:
\begin{equation}
  \Delta_{ij}(t) = (P^t)_{ij} - \pi_j \quad \forall i,j \in \Omega.
\end{equation} From the definition of
\eq{equ:defdelta} for the difference matrix $\Delta$,
together with the normalization of \eq{equ:firstpassagedistri}, we find
\begin{equation}
\pi_j + \Delta_{ij}(t)    =  \sum_{t_1=0}^t F_{ij}(t_1) \glc \pi_j +
\Delta_{jj}(t-t_1) \grc
 = \pi_j \glc 1- \sum_{t¨_1=t+1}^{+\infty} F_{ij}(t_1)\grc +
\sum_{t_1=0}^t
F_{ij}(t_1)  \Delta_{jj}(t-t_1).
\label{deltaetfirst}
\end{equation}
The sum over time $t =0,1,..$ yields the \GUM element $G_{ij} $ of
\eq{equ:GreenDef}
using the normalization of \eq{equ:firstpassagedistri}
and the definition of the \MFPT $\tau_{ij}$ of \eq{equ:mfpt}
as
\begin{align}
G_{ij} & = \sum_{t=0}^{+\infty} \Delta_{ij}(t)
=  \sum_{t=0}^{+\infty} \glb - \pi_j \sum_{t_1=t+1}^{+\infty} F_{ij}(t_1)+
\sum_{t_1=0}^t F_{ij}(t_1)  \Delta_{jj}(t-t_1) \grb
\nonumber \\
& = -  \pi_j \sum_{t_1=1}^{+\infty} F_{ij}(t_1) \sum_{t=0}^{t_1-1}
+ \sum_{t_1=0}^{+\infty}  \sum_{t=t_1}^{+\infty} F_{ij}(t_1)
\Delta_{jj}(t-t_1)
 = - \pi_j \tau_{ij}
+ \underbrace{\sum_{t_1=0}^{+\infty} F_{ij}(t_1)}_{1}
\underbrace{\sum_{t_2=0}^{+\infty}   \Delta_{jj}(t_2)}_{G_{jj}}
\nonumber \\
& = -  \tau_{ij} \pi_j +G_{jj},
\label{equ:greenandfirst}
\end{align}
in other words, \eq{equ:GFirstPassage}.

\section{Mathematical complements}
\label{app:MathDetails}

\subsection{Evaluation of a sum}
\label{sec:SumEvaluation}
We derive the sum $I$ in \eq{equ:KemenyPathGraphSpectrum}. We have
\begin{equation}
I =
\sum_{k=1}^{N-1} \frac{1}{1 - \cosb{k \pi / N}}
= \sum_{k=N+1}^{2N-1} \frac{1}{1 - \cosb{k \pi / N}}.
\end{equation}
With $1 - \cosa{x} = 2 \sinb[2]{x/2}$, it follows that
\begin{equation}
2 I =
\underbrace{\sum_{k=1}^{M-1} \frac{1}{2\sinb[2]{k \pi / M}}}_{(M^2 - 1)/6}
-\half
\end{equation}
with $M=2N$, with the sum computed by Mathematica
(see \app{app:ComputerPrograms}). It follows that
\begin{equation}
 I = \frac{N^2 - 1}{3}.
\label{equ:SumCompute}
\end{equation}

\subsection{\MFPTCAP on the path graph}
\label{app:MFPT_Path}

In this appendix, we derive \eq{equ:tausol}, that is the \MFPT
on the path graph, together with a corresponding equation for different
orderings of argument. For the transition matrix on the path graph
(\eq{equ:PathGraphP}), the \eq{equ:GminusPG} reads
\begin{equation}
\delta_{ij} - \pi_j   =  \sum_{k \in \SET{i\pm 1}} P_{ik} (G_{ij} -  G_{kj} )
= P_{i,i+1} (G_{ij} -  G_{i+1,j} )
+ P_{i,i-1} (G_{ij} -  G_{i-1,j} )\quad \forall i,j \in \SET{1 \TO N},
\label{equ:Gasinversej1d}
\end{equation}
where we set $P_{1,0} = 0$ and $P_{N, N+1} =0$.
Since $j$ is fixed, we may write the  elementary differences as follows:
 \begin{equation}
g^{[j]}_{i+1/2} \equiv G_{i+1,j} - G_{ij} \quad \forall i, j \in \SET{1\TO N}
\label{equ:gdiffj}
 \end{equation}
to rewrite \eq{equ:Gasinversej1d} as the recurrence
\begin{equation}
\delta_{ij} - \pi_j   = - P_{i,i+1} g^{[j]}_{i+1/2}+ P_{i,i-1} g^{[j]}_{i-1/2}
\label{equ:tausolOneDeriv}
\end{equation}

For $i \in \SET{1 \TO j-1}$, \eq{equ:gdiffj} corresponds to the
recursion
\begin{equation}
      g^{[j]}_{i+1/2} = \frac{ P_{i,i-1} g^{[j]}_{i-1/2} +\pi_j }{P_{i,i+1} }
    \label{equ:Gasinversej1dsmaller}
\end{equation}
So for $i=1$ where $P_{1,0}=0$, one obtains the boundary value
\begin{equation}
      g^{[j]}_{3/2} = \frac{ \pi_j }{P_{1,2} }
    \label{equ:Gasinversej1d1}
\end{equation}
and then the solution of the recursion of \eq{equ:Gasinversej1dsmaller}
reads using the detailed-balance property of \eq{equ:DetailedBalancePath}
\begin{equation}
      g^{[j]}_{i+1/2} = \frac{ \pi_j }{ \pi_i P_{i,i+1} } \sum_{k=1}^i \pi_k
\quad \text{for $i \in \SET{1 \TO j-1}$}
  \label{equ:Gasinversej1dsmallersol}
\end{equation}
This then entails, for the sum of the $g^{[j]}_{i+1/2}$ :
\begin{equation}
G_{mj} - G_{jj} = \sum_{i=m}^{j-1}(G_{ij} - G_{i+1,j}) =
    \sum_{i=m}^{j-1}  g^{[j]}_{i+1/2} =
    - \sum_{i=m}^{j-1}
    \frac{ \pi_j }{ \pi_i P_{i,i+1} }  \sum_{k=1}^{i} \pi_k\quad \text{for $m
\in \SET{1 \TO j-1}$}.
  \label{equ:GasinverseNEW1}
\end{equation}

For completeness, we derive \eq{equ:gdiffj} For $i \in \SET{j+1 \TO N}$, it
corresponds to the recursion
\begin{equation}
g^{[j]}_{i-1/2} = \frac{P_{i,i+1} g^{[j]}_{i+1/2} - \pi_j   }{  P_{i,i-1} }.
\label{equ:Gasinversej1dsmallerer}
\end{equation}
The boundary value is obtained for $i=N$,  where $P_{N, N+1}=0$,
\begin{equation}
 g^{[j]}_{N-1/2} = - \frac{ \pi_j   }{  P_{N, N-1} }
  \label{equ:Gasinversej1dbigger}
\end{equation}
and then the solution of the recursion of \eq{equ:Gasinversej1dbigger}
reads using the  detailed balance of \eq{equ:DetailedBalancePath}
\begin{equation}
      g^{[j]}_{i-1/2} = - \frac{ \pi_j }{ \pi_i P_{i,i-1} } \sum_{k=i}^N \pi_k \
\quad \text{for $i \in \SET{j+1 \TO N}$}
\label{equ:Gasinversej1dbiggsol}
\end{equation}
This then entails, for the sum of the $g^{[j]}_{i-1/2}$ :
\begin{equation}
G_{mj} - G_{jj} = \sum_{i=j+1}^{m}(G_{ij} - G_{i-1,j}) =
    \sum_{i=j+1}^{m}  g^{[j]}_{i-1/2} =
    - \sum_{i=j+1}^{m}
    \frac{ \pi_j }{ \pi_i P_{i,i-1} }  \sum_{k=i}^{N} \pi_k\quad \text{for $m
\in \SET{j+1 \TO N}$}.
  \label{equ:GasinverseNEW2}
\end{equation}
The results of \eqtwo{equ:GasinverseNEW1}{equ:GasinverseNEW2}
for the \GM can be translated using \eq{equ:greenandfirst}
  \begin{equation}
  G_{mj} - G_{jj} = -  \tau_{ij} \pi_j
\label{equ:Gmtau}
  \end{equation}
into the following results for the \MFPT $\tau_{mj}$ in the
two regions $m<j$ and $m>j$
 \begin{equation}
 \tau_{mj} =
 \begin{cases}
\sum_{i=m}^{j-1}\frac{ 1 }{ \pi_i P_{i,i+1} } \sum_{k=1}^i \pi_k &
\text{for $m \in \SET{1 \TO j-1}$} \\
\sum_{i=j+1}^{m} \frac{ 1 }{ \pi_i P_{i,i-1} } \sum_{k=i}^N \pi_k &
\text{for $m  \in \SET{j+1 \TO N}$}
\label{equ:tausolAlternative}
 \end{cases}
 \end{equation}
which establishes and extends \eq{equ:tausol}.

\subsection{\MFPTCAP on the lifted path graph}
\label{app:MFPT_LiftedPath}

In this appendix, we derive a general formula that in
a special case yields \eq{equ:tausolLiftedPath}, the
\MFPT on the lifted path graph with transition probabilities as follows:
\begin{equation}
\begin{tikzcd}[
    column sep = 1.5cm,
    row sep = 0.6cm
]
\circled{$\LIFTED{(i-1)}{+}$}
    \arrow{r}{\displaystyle 2 P_{i-1,i}}
&
\circled{\;\;\;$\LIFTED{i}{+}$\;\;\;}
    \arrow[loop above]{}{\underbrace{1-\Omega_i}}
    \arrow{r}{\displaystyle 2 P_{i,i+1}}
    \arrow[xshift=-1.0ex,swap]{d}{\displaystyle \Gamma_i^+}
&
\circled{$\LIFTED{(i+1)}{+}$}
\\
\circled{$\LIFTED{(i-1)}{-}$}
&
\circled{\;\;\;$\LIFTED{i}{-}$\;\;\;}
    \arrow[loop below]{}{\overbrace{1-\Omega_i}}
    \arrow{l}{\displaystyle 2 P_{i,i-1}}
    \arrow[xshift=1.0ex,swap]{u}{\displaystyle \Gamma_i^-}
&
\circled{$\LIFTED{(i+1)}{-}$}
    \arrow{l}{\displaystyle 2 P_{i+1,i}}
\end{tikzcd}
\label{equ:lifted_total_general}
\end{equation}
Here we have two outgoing arrows from any lifted configuration towards other
configurations and only a single one for some lifted configurations at the
edges of the lifted path graph.
We follow the logic of \app{app:MFPT_Path},
which contains the analogous procedure for the path graph.
For the transition matrix of \eq{equ:lifted_total_general},
we refer to sites of the lifted path graph as $(i, \sigma)$ with $i \in
\SET{1\TO N} $ and $\sigma \in \SET{-,+}$. We generalize with respect to
\eq{equ:LiftedTransitionNonReversible} by allowing for a transition probability
$\Gamma_{i}^{\sigma= + }$,
from $(i,+)$ to $(i,-)$,
different from
the transition probability
$\Gamma_{i}^{\sigma=+}$
from $(i,-)$ to $(i,+)$,
and split, as discussed in the main text, the
steady state $\pi_i$ equally between the sites
$(i,+)$ and $(i,-)$, so that $\pi_{i,\sigma} = \half \pi_i$, where $\pi_i$ is
the imposed steady state on the path graph.

We have the
definition that
$P_{ij}=\frac{p}2 \min(1, \frac{\pi_j}{\pi_i})$. We note also that
\eq{equ:lifted_total_general} is a lifting of
\eq{equ:TransitionMatrixGeneralCollapsed} as it  satisfies the lifting
conditions
$\sum_{\sigma }\pi_{i \sigma} = \pi_{i} $, and
$\sum_{\sigma, \sigma'} \pi_{i\sigma} P_{i\sigma, j \sigma'} = \pi_{i}
P_{ij}$, with $\pi_{i\sigma} = \frac12 \pi_i$.

The \GM satisfies, from \eq{equ:GminusPG},
\begin{align}
\delta_{ik} \delta^{\sigma \sigma''}- \frac{\pi_k }{2}
& =  \sum_{j } \sum_{\sigma'}
P_{ij}^{\sigma \sigma'} (G_{ik}^{\sigma \sigma''} -  G_{jk}^{\sigma'
\sigma''} )
\nonumber \\
\intertext{and, taking account of the topology of the lifted path graph, we
have}
& =
  P_{i,i+1}^{\sigma, \sigma}
(G_{ik}^{\sigma \sigma''} - G_{i+1,k}^{\sigma \sigma''} )
+
  P_{i,i-1}^{\sigma, \sigma}
(G_{ik}^{\sigma \sigma''} - G_{i-1,k}^{\sigma \sigma''} )
 +    \Gamma_i^{\sigma} (G_{ik}^{\sigma \sigma''} -
G_{ik}^{-\sigma,\sigma''} )
    \label{equ:Gasinversejsigmadefgeneral}
\end{align}
with the orthogonality relation of \eq{equ:LeftGorthog}
\begin{equation}
0 =   \sum_i \sum_{\sigma} \frac{\pi_i}{2} G_{ik}^{\sigma \sigma''}.
\label{Gorthogjsigma}
\end{equation}
No recursion seems to exist for the \GUM elements for the
general case of \eq{equ:Gasinversejsigmadefgeneral}, which puts into play three
non-vanishing terms starting from a site $(i, \sigma)$ in the bulk and
two non-vanishing terms at site $(1, -1)$ and $(N, 1)$. The skew value
$\Omega_i = p$ for the laziness parameter
is easier, and we treat it in this appendix. In this case,
the counter-rotating term vanishes, $P_{i,i - \sigma}^{\sigma, \sigma} = 1$ and
the recursion can be set up. We perform the corresponding computations here.

The system of \eq{equ:Gasinversejsigmadefgeneral} then reads more explicitly for
$\sigma=+$ and $\sigma=-$ respectively
\begin{align}
\delta_{ik} \delta^{+, \sigma''}- \frac{\pi_k }{2}
& =   2 P_{i,i+1} (\Gcal{+}{i}  -  \Gcal{+}{i+1}  )
 +    \Gamma_i^{+} (\Gcal{+}{i}  -  \Gcal{-}{i} )
 \nonumber \\
\delta_{ik} \delta^{-, \sigma''}- \frac{\pi_k }{2}
& =   2 P_{i,i-1} (\Gcal{-}{i} -  \Gcal{-}{i-1} )
 +    \Gamma_i^{-} (\Gcal{-}{i} -  \Gcal{+}{i}  )
    \label{equ:Gasinversejsigma}
\end{align}

\subsubsection{ First step: \GUM differences on the lifted path graph}

The two elementary differences for the edges on the lifted path graph  are
\begin{align}
\glift{i} & \equiv   \Gcal{-}{i} - \Gcal{ +}{i},
\nonumber \\
\glift{i+\frac{1}{2}} & \equiv   \Gcal{ +}{i+1}  - \Gcal{ -}{i}.
 \label{equ:gskew1}
\end{align}
They satisfy
\begin{align}
\glift{i} + \glift{i+\frac{1}{2}} & =  \Gcal{+}{i+1} - \Gcal{ +}{i}
\nonumber \\
\glift{i -\frac{1}{2}} + \glift{i} & =   \Gcal{ -}{i}   - \Gcal{-}{i-1}
 \label{equ:gskew2}
\end{align}
to rewrite  \eq{equ:Gasinversejsigma} as
\begin{align}
\delta_{ik} \delta^{+, \sigma''}- \frac{\pi_k }{2}
& =   - 2 P_{i,i+1}  \glift{i+\frac{1}{2}}   -    \Omega_i \glift{i}
 \label{equ:gsmall1} \\
\delta_{ik} \delta^{-, \sigma''}- \frac{\pi_k }{2}
& =   2 P_{i,i-1} \glift{i -\frac{1}{2}}  +    \Omega_i \glift{i}
    \label{equ:gsmall2}
\end{align}
in terms of the notation from \eq{equ:lifted_total_general}:
 \begin{equation}
\Omega_i \equiv  2 P_{i,i+1}+ \Gamma_i^{+}  =2 P_{i,i-1}  +  \Gamma_i^{-}.
 \label{equ:omegai}
 \end{equation}
The sum of  \eqtwo{equ:gsmall1}{equ:gsmall2}
yields the closed recursion
\begin{equation}
\delta_{ik} - \pi_k  = 2 P_{i,i-1} \glift{i -\frac{1}{2}}    - 2 P_{i,i+1}
\glift{i+\frac{1}{2}},
    \label{equ:halfinteger}
\end{equation}
for two consecutive half-integer values $\glb i\pm \frac{1}{2} \grb$.

\paragraph{Recursion for $i \in \SET{1 \TO k-1}$:}
For $i \in \SET{1 \TO k-1}$, \eq{equ:halfinteger} corresponds to
the recursion
\begin{equation}
        \glift{i+\frac{1}{2}}  =  \frac{ 2P_{i,i-1}   \glift{i -\frac{1}{2}}
        + \pi_k}{2  P_{i,i+1} }.
    \label{equ:halfintegersmaller}
\end{equation}
For $i=1$, where $P_{1,0}=0$, one obtains the boundary value
\begin{equation}
      \glift{\frac{3}{2}}  = \frac{ \pi_k }{ 2 P_{1,2} }.
    \label{equ:halfinteger1}
\end{equation}
and then the solution of the recursion of \eq{equ:halfintegersmaller}
reads, using the detailed-balance property for $P$,
\begin{equation}
      \glift{i+\frac{1}{2}} = \frac{ \pi_k }{ 2 \pi_i P_{i,i+1} } \sum_{j=1}^i
\pi_j \ \ \ {\rm for } \ \ i \in \SET{1 \TO k-1}.
\label{equ:halfintegersmallersol}
\end{equation}
\Eqq{equ:gsmall1} yields $\glift{i} $ for integer $i$
\begin{equation}
  \glift{i}  =  \frac{ \pi_k   -  4  P_{i,i+1}  \glift{i+\frac{1}{2}}
   }{2\Omega_i}
   = \frac{ \pi_k }{ 2 \Omega_i} \glc 1  - \frac{ 2 }{ \pi_i } \sum_{j=1}^i
\pi_j \grc \ \ \ {\rm for } \ \ i \in \SET{1 \TO k-1}.
\label{equ:Gasinversejsigmasmallintsmall}
\end{equation}
The following sum for $i \in \SET{ \TO k-1}$
is also useful:
\begin{equation}
\glift{i} +  \glift{i+\frac{1}{2}}     =  \frac{\pi_k}{2 \Omega_i }  +\glb 1
-  \frac{ 2  P_{i,i+1}}{\Omega_i}  \grb \glift{i+\frac{1}{2}}
= \frac{\pi_k}{2 \Omega_i }
\glc 1 +
\frac{  (\Omega_i -   2  P_{i,i+1}) }{  \pi_i P_{i,i+1} } \sum_{j=1}^i \pi_j
\grc.
\label{equ:halfintegersmallersolsum}
\end{equation}

\paragraph{Recursion for $i \in \SET{ k+ 1 \TO N}$:}
For $i \in \SET{k+1 \TO N}$, \eq{equ:halfinteger} corresponds
to the recursion
\begin{equation}
  P_{i,i-1}   \glift{i -\frac{1}{2}}  =     P_{i,i+1}  \glift{i+\frac{1}{2}} -
\frac{\pi_k}{2}.
    \label{equ:halfintegerbigger}
\end{equation}
For $i=N$ where $P_{N, N+1}=0$, one obtains the boundary value
\begin{equation}
 \glift{N-1/2}  = - \frac{ \pi_k   }{ 2 P_{N, N-1} },
    \label{equ:halfintegerbiggern}
\end{equation}
and then the recursion of \eq{equ:halfintegerbigger}
reads using the detailed-balance property for $P$:
\begin{equation}
      \glift{i-1/2} = - \frac{ \pi_k }{ 2 \pi_i P_{i,i-1} }
        \sum_{j=i}^N \pi_j \quad \text{for $i \in \SET{k+1 \TO N}$}.
    \label{equ:halfintegerbiggersol}
\end{equation}
\Eqq{equ:gsmall2} yields $\glift{i} $ for integer $i$:
\begin{equation}
  \glift{i}  = - \frac{ \pi_k   + 4 P_{i,i-1}   \glift{i -\frac{1}{2}}
}{2\Omega_i}
  =  \frac{\pi_k  }{ 2 \Omega_i} \glc 1   +    \frac{ 2 }{  \pi_i  }
\sum_{j=i+1}^N \pi_j  \grc
  \quad
    \text{for $i \in \SET{k+1 \TO N}$}.
    \label{equ:Gasinversejsigmasmallintbig}
\end{equation}
The following sum for $i \in \SET{k+1 \TO N} $ is also noteworthy:
\begin{equation}
      \glift{i-1/2} +   \glift{i}=  - \frac{ \pi_k    }{2\Omega_i}
      + \glb 1 -  \frac{  2 P_{i,i-1}    }{\Omega_i} \grb \glift{i -\frac{1}{2}}
      =    - \frac{ \pi_k    }{2\Omega_i} \glc 1
      +  \frac{
      (\Omega_i - 2 P_{i,i-1})}{  \pi_i P_{i,i-1} } \sum_{j=i}^N \pi_j  \grc.
\label{equ:halfintegerbiggersolsum}
\end{equation}
\Eqq{equ:halfinteger} for $i=k$
involves the solution of \eq{equ:halfintegersmallersol} for $i=k-1$
\begin{equation}
\glift{k-\frac{1}{2}} = \frac{ \pi_k }{ 2 \pi_{k-1} P_{k-1,k} } \sum_{j=1}^{k-1}
\pi_j
\label{equ:halfintegersmallersollast}
\end{equation}
and the solution of \eq{equ:halfintegerbiggersol} for $i=k+1$
\begin{equation}
      \glift{k+\frac{1}{2}} =  - \frac{ \pi_k }{ 2 \pi_{k+1} P_{k+1,k} }
\sum_{j=k+1}^N \pi_j
    \label{equ:halfintegerbiggersollast}
\end{equation}
and thus reads, using the detailed-balance property for $P$:
\begin{equation}
1 - \pi_k  = 2 P_{k,k-1} \glift{k -\frac{1}{2}}    -
2 P_{k,k+1}  \glift{k+\frac{1}{2}}
=  \sum_{j=1}^{k-1} \pi_j  \sum_{j=k+1}^N \pi_j,
\label{equ:halfintegercoinciding}
\end{equation}
so that it is automatically satisfied as a consequence of the normalization of
$\pi$.  \Eqq{equ:gsmall1} for $i=k$ that involves
\eq{equ:halfintegerbiggersollast} yields  $\glift{i=k}$:
\begin{equation}
 \glift{k}  = \frac{ \frac{\pi_k }{2}  - \delta^{+, \sigma''}   - 2 P_{k,k+1}
\glift{k+\frac{1}{2}}  }{\Omega_k}
  = \frac{1}{\Omega_k} \glc \frac{\pi_k }{2}  - \delta^{+, \sigma''}   +
\sum_{j=k+1}^N \pi_j  \grc.
\label{equ:gkk}
\end{equation}

\subsubsection{Second step: computing the \GM
$\Gcal{\sigma}{i}$ }

For $i \in \SET{1 \TO  k-1}$, one can use
\eqtwo{equ:gskew2}{equ:halfintegersmallersolsum}
to compute the difference
\begin{align}
    \Gcal{+}{i} -  \Gcal{+}{k}
& = \sum_{m=i}^{k-1}( \Gcal{+}{m} -  \Gcal{+}{m+1})
 = - \sum_{m=i}^{k-1}(g_m + \glift{m+\frac{1}{2}} ),
\nonumber \\
& =  - \sum_{m=i}^{k-1}
\frac{\pi_k}{2 \Omega_m }
\glc 1 +  \frac{  (\Omega_m -   2  P_{m,m+1}) }{  \pi_m P_{m,m+1} }
\sum_{j=1}^m \pi_j \grc.
  \label{equ:gtotdiffsmallerp}
\end{align}
One can then use \eq{equ:gskew1},
\begin{align}
\Gcal{\sigma}{i} &= \Gcal{+}{i} + \delta^{\sigma,-} \glift{i}
\nonumber \\
\Gcal{\sigma''}{k} & = \Gcal{+}{k} + \delta^{\sigma'',-} \glift{k},
  \label{equ:gskewsigma}
\end{align}
to compute from \eqtwo{equ:gskew1}{equ:Gasinversejsigmasmallintsmall} and
as well as \eq{equ:gkk}
the other differences for $i \in \SET{1 \TO k-1}$
\begin{align}
& \Gcal{\sigma}{i}- \Gcal{\sigma''}{k}  =\delta^{\sigma,-}\glift{i}-
\delta^{\sigma'',-}\glift{k} + \Gcal{+}{i}- \Gcal{+}{k}
\nonumber \\
& = \frac{\delta^{\sigma,-} \pi_k }{ 2 \Omega_i} \glc 1  -     \frac{ 2 }{ \pi_i
} \sum_{j=1}^i \pi_j \grc
- \frac{\delta^{\sigma'',-}}{\Omega_k} \glc \frac{\pi_k }{2}    +
\sum_{j=k+1}^N \pi_j  \grc
 - \sum_{m=i}^{k-1}
\frac{\pi_k}{2 \Omega_m }
 \glc 1 +  \frac{  (\Omega_m -   2  P_{m,m+1}) }{  \pi_m P_{m,m+1} }
\sum_{j=1}^m \pi_j \grc.
\label{equ:gtotdiffsmaller}
\end{align}

For $i \in \SET{k+1 \TO N}$, one can use
\eqtwo{equ:gskew2}{equ:halfintegerbiggersolsum}
to compute the difference
\begin{align}
\Gcal{-}{i}- \Gcal{-}{k}
& = \sum_{m=k+1}^{i} (\Gcal{-}{m}- \Gcal{-}{m-1})
 = \sum_{m=k+1}^{i} (\glift{m -\frac{1}{2}} + g_m)
\nonumber \\
& = - \sum_{m=k+1}^{i}  \frac{ \pi_k    }{2\Omega_m} \glc 1
       +  \frac{  (\Omega_m - 2 P_{m,m-1})}{  \pi_m P_{m,m-1} } \sum_{j=m}^N
\pi_j  \grc.
\label{equ:gtotdiffbiggerm}
\end{align}
One can then use \eq{equ:gskewsigma}
to obtain from \eqtwo{equ:gskew2}{equ:Gasinversejsigmasmallintbig}
as well as \eq{equ:gkk} the other differences for $i \in \SET{k+1 \TO N}$
\begin{align}
& \Gcal{\sigma}{i}- \Gcal{\sigma''}{k}  = \delta^{\sigma'',+} \glift{k}  -
\delta^{\sigma,+}\glift{i} +\Gcal{-}{i}- \Gcal{-}{k}
\nonumber \\
& = -  \frac{\delta^{\sigma,+} \pi_k  }{ 2 \Omega_i} \glc 1   +    \frac{ 2 }{
\pi_i  } \sum_{j=i+1}^N \pi_j  \grc
+ \frac{ \delta^{\sigma'',+} }
{\Omega_k} \glc \frac{\pi_k }{2}  - 1   +   \sum_{j=k+1}^N \pi_j  \grc
- \sum_{m=k+1}^{i}  \frac{ \pi_k    }{2\Omega_m} \glc 1
      +  \frac{  (\Omega_m - 2 P_{m,m-1})}
      {  \pi_m P_{m,m-1} } \sum_{j=m}^N \pi_j  \grc
\nonumber \\
& = -  \frac{\delta^{\sigma,+} \pi_k  }{ 2 \Omega_i} \glc 1   +    \frac{ 2 }{
\pi_i  } \sum_{j=i+1}^N \pi_j  \grc
- \frac{ \delta^{\sigma'',+} }
{\Omega_k} \glc \frac{\pi_k }{2}     +   \sum_{j=1}^{k-1} \pi_j  \grc
- \sum_{m=k+1}^{i}  \frac{ \pi_k    }{2\Omega_m} \glc 1
       +  \frac{  (\Omega_m - 2 P_{m,m-1})}{  \pi_m P_{m,m-1} } \sum_{j=m}^N
\pi_j  \grc.
\label{equ:gtotdiffbigger}
\end{align}

At coinciding points $i=k$, the difference between the \GUM elements in
the two copies is given by $\glift{k}$ of \eqtwo{equ:gskew1}{equ:gskew2}
computed in \eq{equ:gkk}:
\begin{align}
 \Gcal{\sigma}{k}- \Gcal{\sigma''}{k}
  &= (\delta^{\sigma,-} \delta^{\sigma'',+}-
\delta^{\sigma,+}\delta^{\sigma'',-}) \glift{k} \\
 &=
 \frac{(\delta^{\sigma,-} \delta^{\sigma'',+}- \delta^{\sigma,+}\delta^{\sigma'',-})}
 {\Omega_k} \glc \frac{\pi_k }{2}  - \delta^{+, \sigma''}   +   \sum_{j=k+1}^N
\pi_j  \grc.
    \label{equ:gtotdiffsmallercoinciding}
\end{align}


\subsubsection{Explicit results for the \MFPT
$\tau^{\sigma \sigma''}_{ik}$ and the \Kemeny $\taukemeny$ }
\label{app:formula_kemeny_lifted}

The \MFPT $\tau^{\sigma \sigma''}_{ik}$ towards site
$(k,\sigma'')$
starting at site $(i,\sigma)$ is given by the difference
of \GUM elements of \eq{equ:greenandfirst}
\begin{equation}
  \tau^{\sigma \sigma''}_{ik} = - \frac{ G^{\sigma \sigma''}_{ik} - G^{\sigma''
\sigma''}_{kk} }{ \frac{\pi_k }{2}}
  =   - \frac{ \Gcal{\sigma }{i} - \Gcal{\sigma'' }{k} }{ \frac{\pi_k
}{2}}.
\label{equ:mftpsigma}
\end{equation}

The results of \eqtwo{equ:gtotdiffsmaller}{equ:gtotdiffbigger}
for the differences of \GUM elements
can thus be translated for the \MFPT of  \eq{equ:mftpsigma}
for $i \in \SET{1 \TO k-1}$:
\begin{equation}
 \tau^{\sigma \sigma''}_{ik}
=   \sum_{m=i}^{k-1}
\frac{1}{ \Omega_m }
\glc 1 +  \frac{  (\Omega_m -   2  P_{m,m+1}) }{  \pi_m P_{m,m+1} }
\Pi_m \grc
 - \frac{\delta^{\sigma,-}  }{  \Omega_i} \glc 1  -     \frac{ 2 }{ \pi_i }
\Pi_i \grc
+ \frac{\delta^{\sigma'',-}}{\Omega_k} \glc 1    + \frac{2}{\pi_k } (1 -
\Pi_{k}) \grc
  \label{equ:mftpsmaller}
\end{equation}
and in the region $i \in \SET{k+1 \TO N}$
\begin{equation}
 \tau^{\sigma \sigma''}_{ik}
=  \sum_{m=k+1}^{i}  \frac{ 1    }{\Omega_m} \glc 1
      +  \frac{  \Omega_m - 2 P_{m,m-1}}{  \pi_m P_{m,m-1} }(1- \Pi_{m-1})
\grc
 +  \frac{\delta^{\sigma,+}   }{  \Omega_i} \glc 1   +    \frac{ 2 }{  \pi_i  }
(1 - \Pi_i)\grc
+ \frac{ \delta^{\sigma'',+} }{\Omega_k} \glc 1    +  \frac{2}{\pi_k }
\Pi_{k-1}  \grc
       \label{equ:mftpbigger}
\end{equation}
while \eq{equ:gtotdiffsmallercoinciding} leads to the \MFPT at coinciding points
$i=k$ in the two different copies $\sigma \ne \sigma''$
  \begin{align}
 \tau^{\sigma \sigma''}_{kk} & = - \frac{2}{\pi_k}
 \glb
 G^{\sigma \sigma''}_{kk} -
G^{\sigma'' \sigma''}_{kk} \grb \\
  &
  = \frac{2}{\pi_k} \frac{(\delta^{\sigma,+}\delta^{\sigma'',-} -
\delta^{\sigma,-} \delta^{\sigma'',+} )}
 {\Omega_k} \glc \frac{\pi_k }{2}  - \delta^{+, \sigma''}   +
\sum_{j=k+1}^N \pi_j  \grc
\label{equ:mftpcoinciding}
\end{align}
that is, more explicitly, for the two non-vanishing cases
 \begin{align}
 \tau^{+-}_{kk} &
   =  \frac{1} {\Omega_k } \glc   1   + \frac{2}{\pi_k }  \sum_{j=k+1}^N \pi_j
\grc,
  \nonumber \\
  \tau^{-+}_{kk} &
   =  \frac{2} {\pi_k \Omega_k } \glc 1 - \frac{\pi_k }{2}  -
\sum_{j=k+1}^N \pi_j  \grc
 =  \frac{2} {\Omega_k \pi_k} \glc  \frac{\pi_k }{2}  +
\sum_{j=1}^{k-1} \pi_j  \grc
 = \frac{1} {\Omega_k} \glc 1  + \frac{2}{\pi_k }  \sum_{j=1}^{k-1} \pi_j,  \grc
\label{equ:mftpcoincidingpm}
\end{align}
so $ \tau^{+-}_{kk} $ corresponds to \eq{equ:mftpsmaller} extrapolated to $i=k$,
while $ \tau^{-+}_{kk} $ corresponds to \eq{equ:mftpbigger} extrapolated to
$i=k$. This establishes and extends \eq{equ:tausolLiftedPath}.

\section{Access to computer programs}
\label{app:ComputerPrograms}

This work is accompanied by the \texttt{LiftedSpectra} software package,
which is published as an open-source project under the GNU GPLv3 license.
The
\texttt{LiftedSpectra} package is available on GitHub as a part of the JeLLyFysh
organization. It contains Python scripts and Mathematica notebooks for the
material discussed in the present paper and, in particular, compares the
numerical
results with exact formulas for the transition-matrix spectra.
The URL of the repository is
\url{https://github.com/jellyfysh/LiftedSpectra.git}.

\end{document}